\documentclass{aa}  
\usepackage{graphicx}
\usepackage{txfonts}
\usepackage{xcolor}
\usepackage{placeins}
\usepackage{multicol}
\usepackage{float}

\usepackage[colorlinks=true,citecolor=blue]{hyperref}
\begin{document} 

    \title{Reference-star differential imaging on SPHERE/IRDIS}
   \author{
        Chen Xie\inst{\ref{lam}}
        \and
        Elodie Choquet\inst{\ref{lam}}
        \and
        Arthur Vigan\inst{\ref{lam}}
        \and
        Faustine Cantalloube\inst{\ref{lam}}
        \and
        Myriam Benisty\inst{\ref{IPAG},\ref{Lagrange}}
        \and
        Anthony Boccaletti\inst{\ref{LESIA}}
        \and
        Mickael Bonnefoy\inst{\ref{IPAG}}
        \and 
        Celia Desgrange\inst{\ref{IPAG},\ref{Heidelberg}}
        \and
        Antonio Garufi\inst{\ref{INAF}}
        \and
        Julien Girard\inst{\ref{STScI}}
        \and
        Janis Hagelberg\inst{\ref{Genève}}
        \and
        Markus Janson\inst{\ref{Stockholm}}
        \and
        Matthew Kenworthy\inst{\ref{leiden}}
        \and
        Anne-Marie Lagrange\inst{\ref{LESIA},\ref{IPAG}}
        \and
        Maud Langlois\inst{\ref{lyon}}
        \and
        François Menard\inst{\ref{IPAG}}
        \and
        Alice Zurlo\inst{\ref{Santiago1},\ref{Santiago2},\ref{lam}}
    }

   \institute{
        Aix Marseille Univ, CNRS, CNES, LAM, Marseille, France \label{lam} \\
        \email{\href{mailto:chen.xie@lam.fr}{chen.xie@lam.fr}}
        \and
        Univ. Grenoble Alpes, CNRS, IPAG, F-38000 Grenoble, France \label{IPAG}
        \and
        Université Côte d’Azur, Observatoire de la Côte d’Azur, CNRS, Laboratoire Lagrange, France \label{Lagrange}
        \and
        LESIA, Observatoire de Paris, Université PSL, CNRS, Sorbonne Université, Université de Paris, 5 place Jules Janssen, 92195 Meudon, France \label{LESIA}
        \and
        Max Planck Institute for Astronomy, Königstuhl 17, D-69117 Heidelberg, Germany \label{Heidelberg}
        \and
        INAF, Osservatorio Astrofisico di Arcetri, Largo Enrico Fermi 5, 50125, Firenze, Italy \label{INAF}
        \and
        Space Telescope Science Institute, 3700 San Martin Drive, Baltimore, MD 21218, USA \label{STScI}
        \and
        Départment d’astronomie de l’Université de Genève,Chemin Pegasi 51, 1290 Versoix, Switzerland \label{Genève}
        \and
        Department of Astronomy, Stockholm University, 10691, Stockholm, Sweden \label{Stockholm}
        \and
        Leiden Observatory, Leiden University, PO Box 9513, 2300 RA Leiden, The Netherlands \label{leiden}
        \and
        CRAL, UMR 5574, CNRS, Université de Lyon, École Normale Supérieure de Lyon, 46 Allée d’Italie, F-69364 Lyon Cedex 07, France \label{lyon}
        \and
        N\'ucleo de Astronom\'ia, Facultad de Ingenier\'ia y Ciencias, Universidad Diego Portales, Av. Ejercito 441, Santiago, Chile \label{Santiago1}
        \and
        Escuela de Ingenier\'ia Industrial, Facultad de Ingenier\'ia y Ciencias, Universidad Diego Portales, Av. Ejercito 441, Santiago, Chile \label{Santiago2}
    }

   \date{Received --; accepted --}

 
  \abstract
   {Reference-star differential imaging (RDI) is a promising technique in high-contrast imaging that is thought to be more sensitive to exoplanets and disks than angular differential imaging (ADI) at short angular separations (i.e., \textless0.3\arcsec). However, it is unknown whether the performance of RDI on ground-based instruments can be improved by using all the archival data to optimize the subtraction of stellar contributions.
   }
   {We characterize the performance of RDI on SPHERE/IRDIS data in direct imaging of exoplanets and disks.}
   {We made use of all the archival data in $H23$ obtained by SPHERE/IRDIS in the past five years to build a master reference library and perform RDI. To avoid biases caused by limited test targets under specific conditions, 32 targets were selected to obtain the average performances of RDI under different conditions, and we compared the performances with those of ADI.}
   {In the point-source detection, RDI can outperform ADI at small angular separations (\textless0.4\arcsec) if the observing conditions are around the median conditions of our master reference library. On average, RDI has a gain of $\sim$0.8 mag over ADI at 0.15\arcsec\,separation for observations under median conditions. We demonstrate that including more reference targets in the master reference library can indeed help to improve the performance of RDI. In disk imaging, RDI can reveal more disk features and provide a more robust recovery of the disk morphology. We resolve 33 disks in total intensity (19 planet-forming disks and 14 debris disks), and 4 of them can only be detected with RDI. Two disks are resolved in scattered light for the first time. Three disks are detected in total intensity for the first time. }
   {RDI is a promising imaging technique for ground-based instruments such as SPHERE. The master reference library we built in this work can be easily implemented into legacy or future SPHERE surveys to perform RDI, achieving better performance than that of ADI. To obtain optimal RDI gains over ADI, we recommend future observations be carried out under seeing conditions of 0.6\arcsec-0.8\arcsec. }


    \titlerunning{RDI performance on SPHERE/IRDIS}
   \maketitle

%
\section{Introduction}

High-contrast imaging is a crucial method for detecting and characterizing wide (\textgreater10~au) giant exoplanets and circumstellar disks around nearby stars. Direct imaging of exoplanets can obtain spectral, orbital, and statistical information of planets to constrain their formation history \citep{Macintosh2015Sci, Nielsen2019, Bowler2020, Vigan2021, Zhang2021Natur}. Direct imaging of disks recovers the disk morphology and surface brightness, which can help to understand the dust grain properties \citep{Milli2017, Chen2020} and potential disk-planet interactions \citep{Ren2020_MWC758}. 

Surveys of exoplanets found that giant exoplanets 1--10 $M_{\rm Jup}$ are rare beyond 30 au, with an occurrence rate of a few percent \citep{Nielsen2019, Vigan2021}. The two largest imaging surveys are the SpHere INfrared Exoplanets (SHINE) project \citep{Desidera2021_SHINE_I}, conducted with Spectro-Polarimetric High-contrast Exoplanet REsearch \citep[SPHERE;][]{Beuzit2019} and the Gemini Planet Imager \citep[GPI;][]{GPI_Macintosh2014PNAS} Exoplanet Survey \citep[GPIES;][]{Macintosh2018}. Each survey targets 500-600 stars. While these surveys have a completeness above 50\% for companions \textgreater10 $M_{\rm Jup}$ at \textgreater10~au, the detection probability drops to 25\% within 5~au. However, radial velocity (RV) studies found a large number of Neptunes and Jupiters at closer orbits (\textless10~au) and a potential peak in the occurrence rate of giant planets at $\sim$3 au \citep{Fernandes2019}, where the water snow line is located. This encourages the development of instruments and techniques to reach deeper detection limits at closer orbits to detect more planets.

The main challenge in the direct imaging of exoplanets and circumstellar disks are the high flux ratios between faint circumstellar objects and bright host stars (i.e., $\sim$$10^{-4}$ arcsec$^{-2}$ for the surface brightness of debris disks and 10$^{-5}$-10$^{-7}$  for giant exoplanets). Dedicated high-contrast imagers can currently suppress the stellar light and reach contrasts of 10$^{-3}$-$10^{-4}$. Then dedicated post-processing methods remove the residual stellar halo and quasi-static features \citep[speckles;][]{Hinkley2007} to detect circumstellar objects.

The majority of the post-processing methods are based on reconstructing proper stellar point spread function (PSF) models from a library of references to subtract stellar contributions. Different techniques have been developed depending on how the references were assembled, either from additional reference stars or from the science data themselves. The reference-star differential imaging (RDI) technique collects stellar references by observing reference or calibration stars with no known circumstellar object \citep{Smith1984Sci, Lafreniere2009}. Assembling stellar references from the science images themselves uses dedicated observing strategies that add diversity to the data to disctinguish the astrophysical signals (planets or disks) from stellar components. 
With respect to the stellar components, the angular differential imaging \citep[ADI;][]{Marois2006} technique uses the azimuthal motion of the astrophysical signal in the temporal direction. Similarly, the spectral differential imaging \citep[SDI;][]{Racine1999} technique uses the radial motion of the stellar components with respect to the astrophysical signal in the spectral direction. In addition, the polarimetric differential imaging \citep[PDI;][]{Kuhn2001} technique uses the fact that scattered light from the dusty disk is polarized, but not that of the starlight, which makes PDI efficient in disk imaging. Because the astrophysical signal can be distinguished from the stellar component, it is possible to model and subtract stellar contributions while retaining the astrophysical signal.

Polarimetric differential imaging is the typical imaging technique for acquiring good contrast for disk detection at small angular separations \citep{Hashimoto2011, Quanz2011, Muto2012}. However, a drawback of PDI is that it can only detect polarized light because any unpolarized flux becomes self-subtracted in differential imaging. Hence, disks with a low polarization fraction can be difficult to detect with PDI. Directly imaging such disks in total intensity is a natural solution. Furthermore, we can derive the polarization fraction from the obtained total and polarized intensities, which is an important quantity for understanding the dust properties in the disk. 

Angular differential imaging is the primary imaging technique for the SHINE and GPIES surveys in searching for exoplanets and imaging disks in total intensity. However, ADI has limitations at small angular separations (e.g., \textless0.3\arcsec~on 8m class telescopes) mainly due to the self-subtraction effect \citep{Marois2006}. The astrophysical signal (i.e., planets) needs to move a certain arc-length to be distinguishable from stellar components to avoid subtracting itself, which is the so-called self-subtraction effect. For details of ADI limitations (e.g., timing constraints, limited sky coverage, limited sensitivity around the inner working angle (IWA), or sharping azimuthal features of extended sources), we refer to the introduction of \cite{Ruane2019}. As a result, the self-subtraction effect degrades the sensitivity of ADI in detecting exoplanets at small angular separations and limits the accurate recovery of disk morphology \citep{Milli2012}. Therefore, the sensitivity of current instruments using ADI at \textless0.3\arcsec is limited to massive objects and does not include the bulk of lower-mass planets expected from models or RV studies. A similar effect also limits the performance of SDI on current instruments because it also uses the science data themselves to build the stellar model.

To avoid the self-subtraction effect and improve the sensitivity at short separations (\textless0.3\arcsec), we can use RDI as an alternative technique for the direct imaging of exoplanets and disks in total intensity. The key aspect of RDI is to assemble proper stellar references. There are different strategies to assemble references, which can be divided into two types. The first strategy type is to collect stellar references similar to science images, which can be achieved by 1) observing binaries \citep{Kasper2007, BDI_Rodigas2015}, 2) using the stars observed under similar conditions \citep{YSES_Bohn2020MNRAS} or at the same night \citep{Weinberger1999, Xuan2018}, and 3) observing science and reference stars nearly simultaneously \citep{Wahhaj2021}. The second strategy type is to use all the archival data obtained with a stable instrument and then down-select those that are most similar to the science data \citep{Lafreniere2009, Soummer2011, ALICE_Choquet2014SPIE}.

Space-based instruments usually have more stable PSFs than ground-based instruments. Hence, RDI was successfully applied to space-based instruments \citep[e.g.,][]{Schneider2014, Schneider2016, Schneider2018} and became their primary imaging technique, including \textsl{Hubble Space Telescope} (\textsl{HST})/NICMOS and \textsl{HST}/STIS \citep{ALICE_Choquet2014SPIE, Hagan2018, Ren2021}. Recently, \cite{Sanghi2022} compared the performances of RDI and ADI based on the \textsl{HST}/WFC3 observation of \object{PDS\,70} \citep{Zhou2021}. They found that the performance of RDI is worse than that of ADI, which might be attributed to the lack of a high-quality reference library.

In ground-based instruments, RDI is not commonly used because of the rapid atmospheric turbulence. The recent developments of RDI have achieved some gains at short angular separations with ground-based instruments equipped with adaptive optics (AO) system, such as Keck/NIRC2 \citep{Xuan2018, Ruane2019} and the star-hopping mode on SPHERE \citep{Wahhaj2021}. Both studies adopted the first type of strategy that collects stellar references similar to science images. It remains unknown whether the performance of RDI can be improved by using all the archival data to optimize the subtraction of stellar contributions. \cite{Gerard2016} assembled a reference library of 207 reference targets from GPIES and applied RDI on GPI data of \object{51\,Eri}. However, no significant improvement of RDI over ADI was found for \object{51\,Eri} data. Nevertheless, it is worth exploring whether the RDI performance improves with the number of reference targets. More importantly, the study was established on a single test case (51~Eri), which was observed under specific observing conditions with specific observation settings. It is possible that RDI may achieve different performances for observations obtained in different observing and instrument conditions. More studies are needed to avoid potential bias caused by limited test cases to obtain a general performance of RDI on ground-based instruments. %

In this paper, we present a thorough analysis of the RDI performance on SPHERE using all the archival $H23$ data observed in the past five years, including $\sim$$2.6\times10^{5}$ images from more than 1000 observations. We explore the RDI performance in the detection of point sources and disk imaging. In Sect.~\ref{sec:method} we describe the data used in this work and our RDI and ADI approaches. In Sect.~\ref{sec:point_srouce_detection} we present the RDI performances in the detection of point sources under different conditions and compare them with that of ADI. In Sect.~\ref{sect:disk_imaging} we present the advantages of RDI in disk imaging and show the disks detected by SPHERE in total intensity. Finally, we give the conclusions and discuss our results in Sect.~\ref{sec:conclusions}.

\section{Methods}
\label{sec:method}

\subsection{Data}
The SPHERE instrument \citep{Beuzit2019} on the Very Large Telescope (VLT) has delivered high-quality images since first light in May 2014. SPHERE has three science instruments: the infrared dual-band imager and spectrograph \citep[IRDIS;][]{IRDIS_Dohlen2008SPIE}, the integral field spectrograph \citep[IFS;][]{IFS_Claudi2008SPIE}, and the Zurich imaging polarimeter \citep[ZIMPOL;][]{ZIMPOL_Schmid2018A&A}. Both IFS and IRDIS work in the near-infrared and can provide high Strehl ratios (e.g., \textgreater75\% in $H$), while ZIMPOL works in the visible and provides lower Strehl ratios (e.g., \textless50\% in $R$). IRDIS is a dual-band imager \citep[DBI;][]{Vigan2010} that produces simultaneous images at two nearby wavelengths that were selected around expected features in the spectrum of young giant exoplanets. Both IFS ($Y$-$J$) and IRDIS ($H$) have a similar spectral resolution of $\sim$30, but IFS provides 39 spectral channels. 

To quantify the performance of RDI on ground-based high-contrast imagers such as SPHERE, we used IRDIS due to its high Strehl ratio in the infrared and similar spectral resolution as IFS, but is less complex than IFS data. The diversity of the reference images can be used to build the proper stellar model only with a large number of reference images. The most commonly used coronagraph is the apodized pupil Lyot coronagraph in its $\texttt{N\_ALC\_YJH\_S}$ configuration \citep{Carbillet2011, Guerri2011}, with a mask diameter of 185~mas. The three most commonly used filter pairs are  $\texttt{DB\_H23}$, $\texttt{DB\_K12}$, and $\texttt{BB\_H}$, with $\sim$$1.3 \times 10^{5}$, $\sim$$8.7 \times 10^{4}$, and $\sim$$5.8 \times 10^{4}$ exposures in the public archive as of 2021 January 1, respectively. 
To achieve optimal performance, we adopted $\texttt{DB\_H23}$ in this work because this filter pair has the highest number of exposures. The total number of coronagraphic images from two bands (dual-band mode; $H2$:1.593~$\mu$m and $H3$:1.667~$\mu$m) is about $2.6 \times 10^{5}$.

The raw data were processed using the $\texttt{vlt-sphere}$\footnote{\url{https://github.com/avigan/SPHERE}, version 1.4.2} pipeline \citep{SPHERE_pipeline_Vigan2020} to produce calibrated data cubes and stellar PSF images without the coronagraph mask. SPHERE real-time computer data (also called SPARTA files) were also processed by the $\texttt{vlt-sphere}$ pipeline to provide the observing conditions used in the performance analyses. Because real-time computer data and science exposures were not synchronous, we associated each science image with the observing condition that was closest in time. 

\subsection{Reference-star differential imaging}
\label{subsec:RDI}

\subsubsection{Construction of the master reference library}
\label{subsec:building_ref_lib}
Building the master reference library is the key element in RDI. This was done in two steps, the image alignment and the identification of poor reference stars. Throughout the paper, the master reference library refers to all the available reference images, and the reference library refers to a selected subset of the master reference library for the subtraction of stellar contributions in the individual science image. The selection of a subset of the master reference library is described in Sect.~\ref{subsection:image_select_psf_sub}. 

The image alignment is important because the diffraction features and speckles must be aligned in all the images and to optimize the subtraction of stellar contributions. Furthermore, it can maximize the signal of any astrophysical object in the combined image. SPHERE typically uses satellite spots in the first and last images of an observation to locate the star center behind the coronagraph. During the entire science observation, SPHERE relies on the differential tip-tilt sensor control to maintain the star at the same position behind the coronagraphic mask. However, the differential tip-tilt sensor loop runs at 1 Hz, so that some residual jitter of the images can occur at a faster rate. This may therefore induce some small shift. We discuss the pointing stability of SPHERE/IRDIS in Appendix~\ref{subsec:pointing_stability}.

The first step of the image alignment is to choose a reference template image to which all the images will be aligned. A coronagraphic image with a high signal-to-noise ratio (S/N) can be a good reference template to improve the accuracy of the image alignment. We arbitrarily adopted the bright star (\citealp[$H$$\sim$3.7 mag;][]{2MASS_Skrutskie2006}) \object{HD\,121156} (program 098.C-0583, PI: Pantoja), which was observed on 2017 Febuary 5 under excellent conditions (seeing: 0.33\arcsec, $\tau_{0}$:16.89 ms, and Strehl:93\%) with a clear correction ring (see, Fig.~\ref{Fig:frame_registration}). The correction ring marks the boundary of the AO-corrected area (inside) and the seeing-limited area (outside). For the \texttt{DB\_H23} filter pair, the control radius (inner edge of the correction ring) is about 0.7\arcsec-0.8\arcsec. An image mask (see the magenta circles in Fig.~\ref{Fig:frame_registration}) was used to only focus on the correction ring in the image alignment. Before performing the image alignment, we removed the low spatial frequencies of the stellar halo \citep[e.g., wind-driven halo; ][]{Cantalloube2020} using a median filter with a size of 15 $\times$ 15 pixels. The stellar halo was only removed with the median filter in the image alignment to obtain the offsets, not to perform the subtraction of stellar contributions.

Each image was only aligned to the selected reference template by minimizing the loss function $L$ as 
\begin{equation}
\label{equ:loss_function}
\arg \min_{a,\  x,\  y} L \  =\  \log \left( \sum^{N_{\text{pix} }}_{i=1} \left( \left( a\  I_{i}\left( x,y\right)  \  -\  T_{i}\right)  M_{i}\right)^{2}  \right),   
\end{equation}
where $a$, $x$, $y$, $i$, and $N_{\rm pix}$ are the intensity scaling factor, the offset on the x-axis, the offset on the y-axis, the image pixel index, and the total number of pixels, respectively. $T$, $M$, and $I(x,y)$ are the reference template, the image mask, and a given image after being shifted by given offsets on the x- and the y-axis, respectively. 

All the images in our database were aligned to the same reference template of the same band after being shifted by derived offsets. $H2$ and $H3$ images were processed separately. 
To ensure that the images in different bands were also aligned, we aligned the reference templates in $H2$ and $H3$ beforehand. This was done by scaling the reference template in $H3$ according to its wavelength and aligning it with the template in $H2$. 
Throughout the paper, we excluded the images that failed in the image alignment, which could be caused by bright sources around the correction ring, failed coronagraphic images (i.e., star outside the coronagraph mask or fail AO corrections), short exposures (shorter than a few seconds), and poor observing conditions. About 12\% of the images failed in the image alignment (see Appendix~\ref{appendix:frame_registration} for details).

Then, as a second step, we analyzed each target in order to identify poor reference stars. Any reference image that contained point sources and/or extended sources was considered to be a poor reference in RDI, which results in an incorrect starlight subtraction \citep{ALICE_Choquet2014SPIE}. After the image alignment, we performed the subtraction of stellar contributions using ADI and then RDI (see, Sects.~\ref{subsec:ADI} and~\ref{subsec:RDI} for the details) to exclude poor reference stars by visual inspections. Most of the points and extended sources can be identified after ADI. Then we performed our RDI using the master reference library obtained after ADI to further remove poor references. After the image alignment and the identification of bad reference stars, we assembled the master reference library, which contains about  $7\times10^{4}$ images per band from 725 observations. Throughout the paper, we define an observation as a complete observing sequence of a target that consisted of successive frames observed in the pupil-stabilized mode. A given target may contain multiple observations obtained at different nights under different observing conditions.

\subsubsection{Image selection and subtraction of stellar contributions}
\label{subsection:image_select_psf_sub}
Some level of image selection is necessary to find the optimal reference images that can effectively improve the subtraction performance of stellar contributions \citep{Ruane2019}. It is not necessary to use the whole master library because it contains reference images that were obtained in very different observing conditions that are significantly different from a given science image. The best two methods tested in \cite{Ruane2019} are the mean square error (MSE) and structural similarity index metric (SSIM). MSE and SSIM provided similar correlations and performances. However, MSE required fewer computation resources, which is suitable for applying it to a very large number of images.

In the image selection, we first normalized both science and reference images with a robust scaler\footnote{The robust scaler was defined as $I_{\rm scaled} = (I_i - {\rm med}(I))/(Q_3(I) - Q_1(I))$, where $I$, $i$, ${\rm med}(I)$, $Q_3$, and $Q_1$ are the image, the pixel index, the sample median, the third quartile, and the first quartile, respectively.} that removed the median and scaled the data according to the interquartile range. Then we computed the MSE values between the science image and each of the frame in the master reference library, which was given by
\begin{equation}
\label{equ:MSE}
\text{MSE}^{\left( k\right)} \  =\  \frac{1}{N_{\text{pix} }} \sum^{N_{\text{pix} }}_{i=1} \left( R^{\left( k\right)}_{i}-S_{i}\right)^{2},
\end{equation}
where $k$, $N_{\rm pix}$, $R$, and $S$ are the frame index in the master reference library, the number of pixels, a reference image, and a given science image, respectively. An image mask was used with an inner mask of 8 pixels ($\sim$0.1\arcsec) in radius and an outer mask of 60 pixels (0.735\arcsec) in radius. The size of the outer image mask was adopted to only focus on the AO-corrected area.

For each science image, we selected a reference library to reconstruct a stellar coronagraphic image using principal component analysis \citep[PCA;][]{Soummer2012}. 
Multiple sizes of reference libraries and numbers of subtracted principal components (PCs) were used in the main analyses, such as making contrast curves (see the main text in Sect.~\ref{subsec:contrast_curve}) and disk imaging (see the main text in Sect.~\ref{sec:disk_imaging}). To identify poor reference stars, we used 500 and 2000 images as the sizes of the reference library to perform RDI. The number of PCs to be subtracted are 2\%, 5\%, 10\%, 15\%, 20\%, and 50\% of the reference library.  
After the subtraction of stellar contributions, the residual science cubes were derotated and mean combined to form the residual images in $H2$ and $H3$ for each science target. We also created a final residual image by combining the two residual images in $H2$ and $H3$.

\subsection{Angular differential imaging}
\label{subsec:ADI}
We systematically processed all the targets after the image alignment using ADI. The aims were to exclude the poor reference stars from the master reference library and  provide a standard for the comparison with RDI. 
After the image alignment (see, Sect.\ref{subsec:building_ref_lib}), a stellar coronagraphic image was built for each science image using PCA. In the identification of poor reference stars, the number of PCs to be subtracted were 2\%, 5\%, 10\%, 15\%, 20\%, and 50\% of the total number of science images. Then, each science image was subtracted by the corresponding stellar image to remove the stellar contributions. $H2$ and $H3$ bands were processed separately. Similar to RDI, residual cubes were derotated and mean combined to form the residual images in $H2$ and $H3$ bands. A final residual image was also obtained by combining two residual images in $H2$ and $H3$.

\subsection{Contrast curves}
\label{subsec:contrast_curve}
In the detection of point sources, the performances of RDI and ADI can be quantified and compared via their contrast curves. To calculate the 5$\sigma$ contrast curve $C$, we followed
\begin{equation}
\label{equ:contrast}
C=\left( 5t_{\rm{s} tu}\sigma +f_{\rm{r} es}\right)  T^{-1}_{\rm{i} ns}f^{-1}_{\rm{s} tar},
\end{equation}
where $t_{\rm stu}$, $\sigma$, $f_{\rm res}$, $T_{\rm ins}$, and $f_{\rm star}$ are the correction factor for small sample statistics \citep{Mawet2014}, noise, the residual flux after the PCA subtraction, the throughput of the PCA subtraction, and the stellar flux observed without the coronagraph, respectively. All terms are radial dependent from the star center and sampled by increasing the radial separation in steps of one full width at half maximum (FWHM), except for the stellar flux.   
The noise and residual flux were estimated in the residual image as fluxes integrated within the resolution elements. To do this, we sampled each annulus at a given radial separation with apertures with a diameter of 1~FWHM. The residual flux and the noise were estimated as the mean and standard deviation of the fluxes integrated within these apertures, respectively. By fitting the stellar PSF observed by SPHERE without the coronagraph mask, we obtained the size of the aperture (1~FWHM) and the stellar flux. 

The throughput of the PCA subtraction was estimated by injecting a simulated planet. The artificial planet was created based on the stellar PSF with a flux scaled to 20 times the noise. We show in Appendix~\ref{appendix:Validation_throughput_estimation} that the level of injected flux has a negligible impact on the estimation of the throughput. To reduce the computation time, eight artificial planets were injected each time with a radial separation of 2 FWHM and azimuthal separation of 90$^{\circ}$ to avoid biasing the throughput estimation. The throughput at injected locations was estimated via the ratio of recovered and injected flux. We used eight different position angles (each separated by 45$^{\circ}$) to cover the field of view. Finally, we sampled each radial separation with eight spatial locations and obtained an azimuthally averaged $C$.

A few parameters affect the performances of RDI and ADI, such as the number of subtracted PCs and the size of the reference library. To explore the parameter space, a grid of parameters was adopted for RDI with sizes of the reference library of 200, 500, 1000, 2000, 3000, 5000, and 10000 images and numbers of PCs of 2\%, 5\%, 10\%, 15\%, 20\%, 50\%, and 70\% of the reference library. We used the same parameters for ADI, except for the size of the reference library, which was the number of science images for each observation. Both RDI and ADI use the same approach to estimate the contrast curve for the detection of point sources. The only difference is the dataset that was used in PCA to calculate PCs.

We finally adopted the best contrast value that we achieved with our grid of parameters at each angular separation to form the optimal contrast curve. The concept of optimal contrast was used by \cite{Xuan2018} to describe the best achievable contrast at a given angular separation. Throughout the paper, we use the optimal contrast curve as the contrast curve we achieved for each target.

\subsection{Sample selections}
\label{subsec:sample_selection}

  \begin{figure*}
  \centering
    \includegraphics[width=0.98\textwidth]{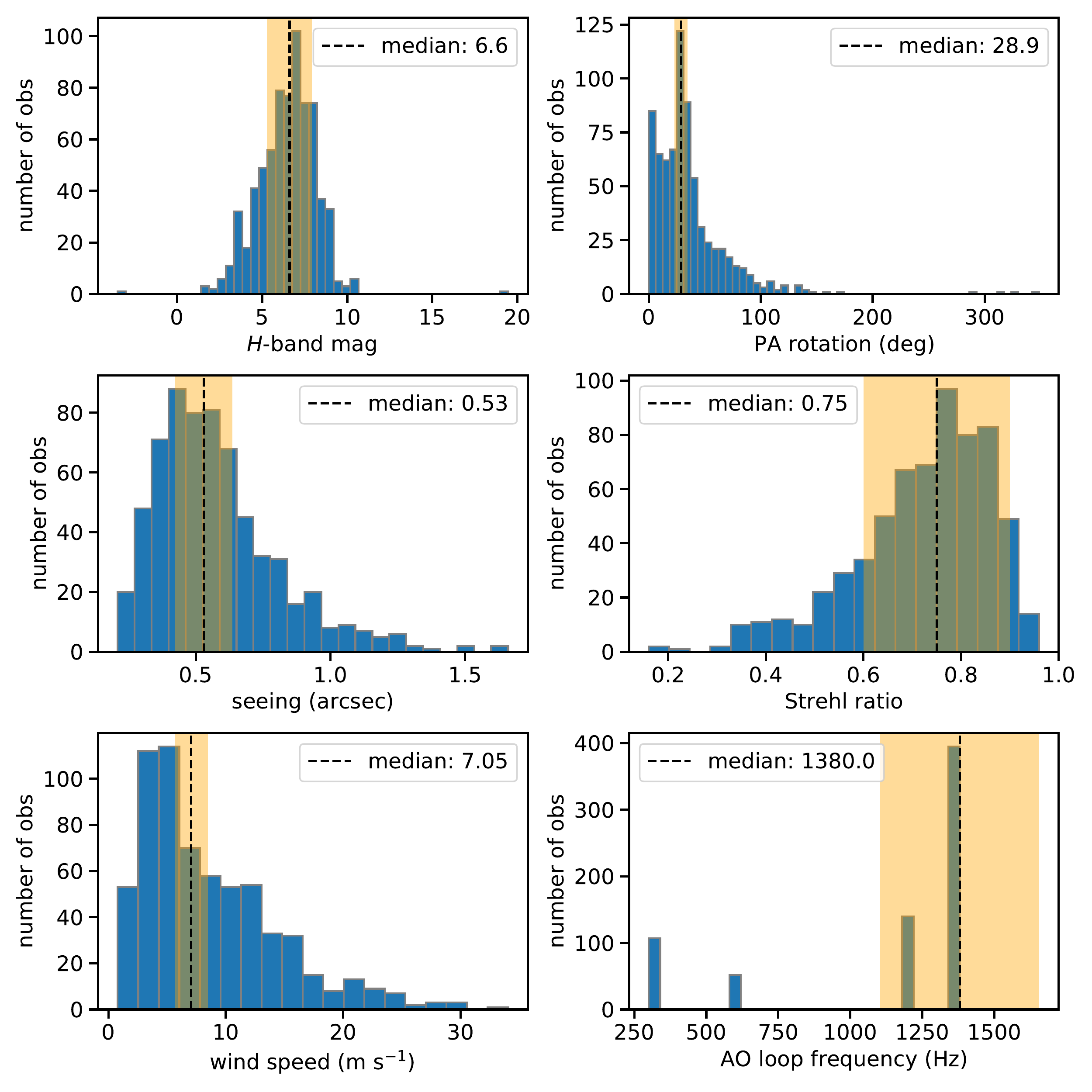}
      \caption{Properties of all 725 observations in our master reference library. \textsl{From top left to bottom right panel:} distributions of $H$-band magnitude, PA rotation, seeing, Strehl ratio, wind speed, and AO loop frequency. The corresponding median values (vertical dashed lines) of $H$ magnitude, PA rotation, seeing, Strehl ratio, wind speed, and AO loop frequency are 6.6~mag, 28.9$^{\circ}$, 0.53\arcsec, 0.75, 7.05~m~s$^{-1}$, and 1380 Hz. The yellow shaded region represents 20\% of the corresponding median value. The conditions of seeing, Strehl ratio, and wind speed came from the SPARTA files.
              }
      \label{Fig:distribution_ob_condition}     
  \end{figure*}

The aim of this paper is to illustrate the general performance of RDI using the diversity of SPHERE/IRDIS data and compare it with that of ADI. The atmospheric conditions have strong impacts on the performance of RDI and ADI. Furthermore, ADI is affected by the self-subtraction effect due to the limited parallactic angle (PA) rotations of the science data. Therefore, we specifically study these two effects independently in Sects.\ref{subsec:RDI_vs_seeing} and \ref{subsec:RDI_vs_PA}. To obtain general RDI performances for different observing and instrument conditions, we selected two groups of targets from the master reference library. The selected targets are free from the contamination of celestial sources, which may otherwise bias the performance analysis.

Ideally, the same observing conditions should result in the same stellar halo and speckle patterns if the instrument were perfectly stable. Therefore, RDI observations under similar observing conditions may lead to a larger number of well-matching reference images, which was the idea for multiple surveys designed to use RDI \citep[e.g.,][]{YSES_Bohn2020MNRAS}. To explore the impact of the observing conditions, we adopted seeing condition bins of 0.2\arcsec, 0.4\arcsec, 0.6\arcsec, 0.8\arcsec, and 1.0\arcsec. Only fewer than 5\% of observations in our master reference library have a seeing higher than 1.0\arcsec. In addition to the seeing condition, we adopted the field rotation, wind speed, Strehl ratio, $H$ band magnitude, and AO loop frequency as control variables. We selected targets with values within $m \pm 0.2m$, where $m$ is the median value of the control variable. Here we adopted an arbitrary fraction of 20\%, which is a trade-off between having similar conditions and obtaining enough targets in each bin to perform a minimalist statistical analysis. The distributions of observing conditions in our master reference library and the corresponding median values are shown in Fig.~\ref{Fig:distribution_ob_condition}. The selected sample of targets with different seeing conditions is listed in Table~\ref{table:sample_delta_Seeing}.

The main limitation of ADI is the self-subtraction effect that degrades the sensitivity of point-source detection at small angular separations (i.e., \textless0.3\arcsec) and modifies any extended structure \citep{Marois2006, Milli2012}. For example, a required PA rotation is $\sim$28$^{\circ}$ at a separation of 0.1\arcsec~in order to move the PSF of a point source by an arc-length of 1~FWHM (50~mas with SPHERE).  However, our master reference library of over 700 observations has a median PA rotation of $\sim$29$^{\circ}$. A significant fraction of SPHERE/IRDIS data suffers the self-subtraction effect at small angular separations. In contrast, RDI does not have the self-subtraction effect because it builds stellar models from references that have been vetted to be free from companions. We adopted PA rotation bins of 5$^{\circ}$, 20$^{\circ}$, 30$^{\circ}$, 40$^{\circ}$, 60$^{\circ}$, and 80$^{\circ}$ to study its impact on the performance. Only fewer than 8\% of observations in our master reference library have PA rotations higher than 80$^{\circ}$. The rest of the conditions (seeing, wind speed, Strehl ratio, H band magnitude, and AO loop frequency) are about the median values within 20\% of their median values. The selected sample of targets with different PA rotations is listed in Table~\ref{table:sample_delta_PA}. Each PA bin contains four to five targets, except for the last bin of 60$^{\circ}$ - 80$^{\circ}$ , which has only two targets.

\section{RDI performance for point-source detection}
\label{sec:point_srouce_detection}

\subsection{RDI performance as a function of seeing conditions}
\label{subsec:RDI_vs_seeing}

  \begin{figure*}
  \centering
    \includegraphics[width=0.98\textwidth]{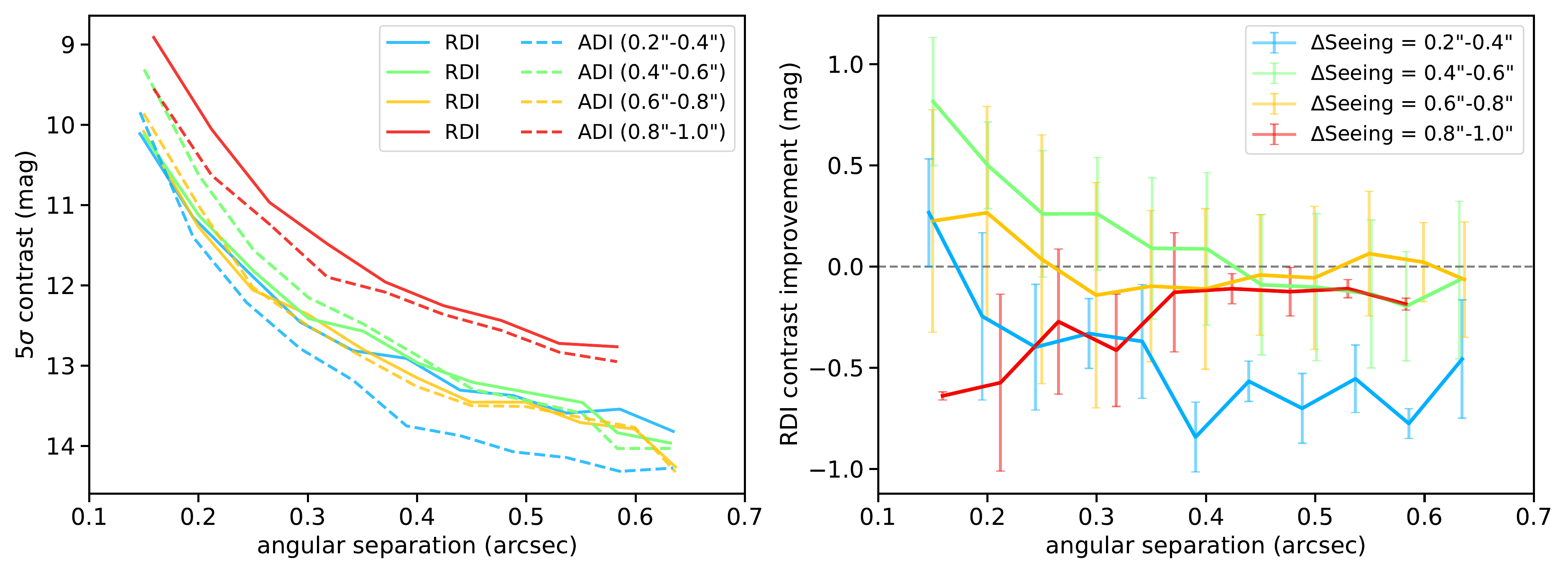}
      \caption{Comparison of RDI and ADI contrasts. \textsl{Left panel:} Contrast curves (5$\sigma$) for the selected sample with the RDI (solid lines) or ADI (dashed lines) reduction, averaged over targets in each bin of seeing conditions. \textsl{Right panel:} RDI contrast improvement over ADI. The error bar represents the scatter of the RDI improvements of the targets in each bin. Four bins of seeing conditions were used.
              }
         \label{Fig:RDI_vs_ADI_delta_Seeing}
  \end{figure*}

  \begin{figure*}
  \centering
    \includegraphics[width=0.98\textwidth]{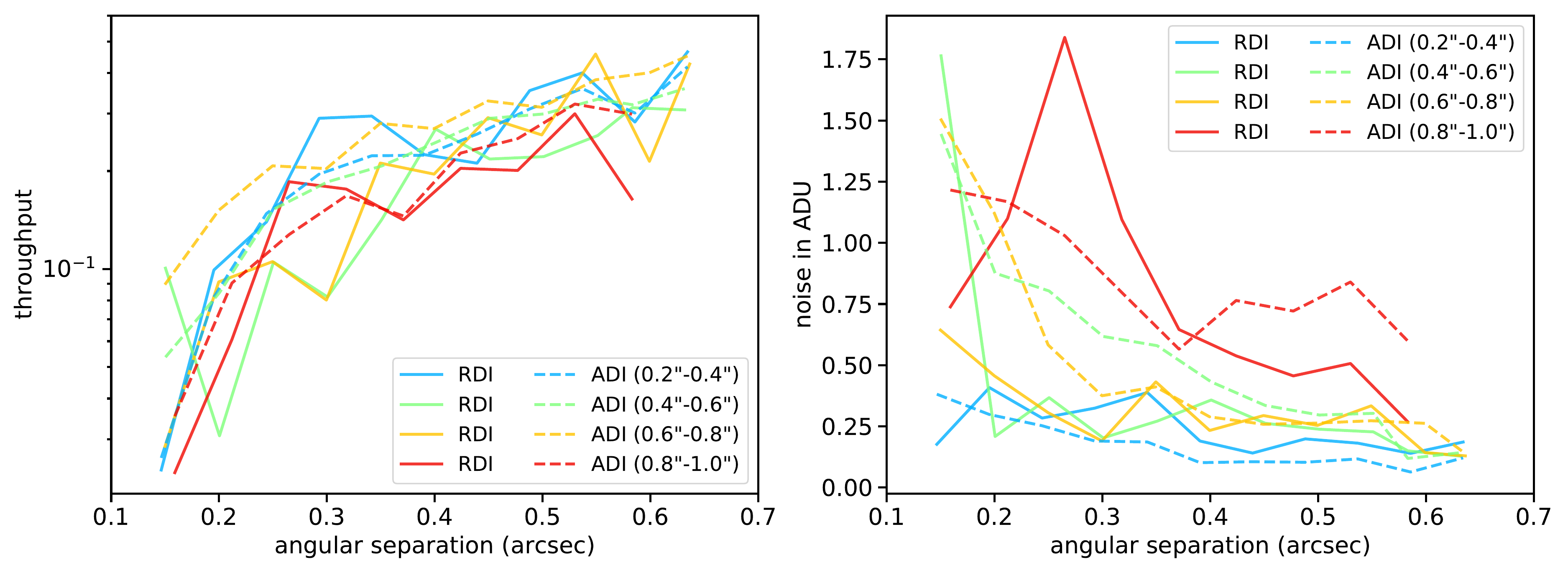}
      \caption{The throughput and noise correspond to the contrast curve shown in Fig.\ref{Fig:RDI_vs_ADI_delta_Seeing}. \textsl{Left panel:} Throughput of PCA subtraction using RDI (solid lines) or ADI (dashed lines) as a function of angular separations. \textsl{Right panel:} Noise in the residual image processed with RDI (solid lines) or ADI (dashed lines) as a function of angular separations. Four bins of seeing conditions were used. The throughput and noise were averaged over targets in each bin of seeing conditions. 
              }
         \label{Fig:RDI_vs_ADI_delta_Seeing_throughput_noise}
  \end{figure*}

We selected 19 targets out of 725 observations in our master reference library based on the criteria described in Sect. 2.5, listed in Table~\ref{table:sample_delta_Seeing}. Each observation has similar observing conditions (i.e., PA rotation, wind speed, Strehl ratio, $H$-band magnitude, and AO loop frequency) that are about the median values shown in Fig.~\ref{Fig:distribution_ob_condition}. Only the seeing conditions are significantly different and in a range of 0.2\arcsec-1.0\arcsec. 
We estimated the contrast curves of RDI and ADI on selected targets (see Sect.~\ref{subsec:contrast_curve} for the details). In each bin of the seeing conditions, we took the mean of the contrast curves of the test targets. We estimated the uncertainty by taking the standard deviation of the results for the targets in each bin, representing the scatter of RDI gains over ADI.

Fig.~\ref{Fig:RDI_vs_ADI_delta_Seeing} shows the contrast curves of RDI and ADI as a function of angular separations in each seeing bin. The RDI and ADI contrasts improve with better seeing conditions, which is expected because a better AO performance can be achieved with better seeing. The improvement of RDI over ADI is also shown in Fig.~\ref{Fig:RDI_vs_ADI_delta_Seeing}. RDI outperforms ADI at small separations (\textless 0.4\arcsec) for observations under seeing conditions of about 0.4\arcsec-0.6\arcsec. These seeing conditions are about the median value of our master reference library, which is 0.53\arcsec\,(see, Fig.~\ref{Fig:distribution_ob_condition}). At a separation of 0.15\arcsec, RDI reaches its peak efficiency with an average gain of $0.8 \pm 0.3$ mag over ADI for observations with seeing conditions ranging from 0.4\arcsec-0.6\arcsec. In slightly worse seeing conditions of 0.6\arcsec-0.8\arcsec, RDI and ADI have a similar performance. In excellent seeing conditions of \textless0.4\arcsec, ADI is very efficient at large separations, but RDI still outperforms ADI at 0.15\arcsec\,separation. However, ADI outperforms RDI for observations under extreme conditions (i.e., 0.2\arcsec-0.4\arcsec~or 0.8\arcsec-1.0\arcsec) for most separations. 

The degradation of RDI performance in extreme seeing conditions can be explained by the fact that there are fewer well-matching references for observations under extreme conditions. A larger number of well-matching reference images can only be found when the observing conditions are similar (i.e., about median values in Fig.~\ref{Fig:distribution_ob_condition}), which leads to better PCA subtractions and deeper contrasts. Consequently, increasing the size of the master library reference over time will statistically increase the number of matching reference images for all conditions and improve the overall performance of RDI. We further explore the impact of the size of the master reference library in Sect.\ref{subsect:impact_of_m_size}.

The uncertainty shown in Fig.~\ref{Fig:RDI_vs_ADI_delta_Seeing} is affected by multiple factors. The small sample size directly affects the statistic. The scatter of observing conditions among the targets may also lead to scatter in RDI gains over ADI. 
Therefore, we selected targets with similar observing conditions. However, we need to balance being selective to have targets under similar conditions and having enough targets in each bin. 
Overall, the scatter in RDI gains over ADI justifies the necessity of selecting a sample to generate an average performance for certain observing conditions. Evaluating the RDI performance based on only a few targets under different conditions will lead to a biased result.

Fig.~\ref{Fig:RDI_vs_ADI_delta_Seeing_throughput_noise} shows the throughput of the PCA subtraction and noise in the residual image after the processes of RDI and ADI (see Sect.~\ref{subsec:contrast_curve} for the details), which correspond to the contrasts shown in Fig.~\ref{Fig:RDI_vs_ADI_delta_Seeing}. The differences in throughput and noise after the reductions of RDI and ADI are also shown in Fig.~\ref{Fig:comparison_TT_NN}. Noise in the RDI and ADI images decreases with the increase in angular separations. Higher noise at short separations for both RDI and ADI is expected because both speckle noise and photon noise are stronger at short separations. Moreover, better conditions lead to a better AO performance, resulting in less noise for all separations. Both RDI and ADI throughputs increase from about 5\% to 50\% with angular separations for all bins of seeing conditions. 

The RDI technique was expected to have higher throughput than the ADI technique. However, completely removing the speckle noise requires RDI to subtract a large number of PCs, which may lead to some levels of oversubtraction and hence lower its throughput. Therefore, the RDI throughput is not higher than that of ADI in Fig.~\ref{Fig:RDI_vs_ADI_delta_Seeing_throughput_noise} (see also Fig.~\ref{Fig:comparison_TT_NN}).  As a comparison, the self-subtraction effect in ADI not only lowers the throughput, but also contributes to removing the speckles, thus lowering the noise. Overall, the optimal contrast was the balance between higher throughput and lower noise, as shown in Eq.~\ref{equ:contrast}.

\subsection{RDI performance as a function of PA rotations}
\label{subsec:RDI_vs_PA}

 \begin{figure*}
  \centering
    \includegraphics[width=0.98\textwidth]{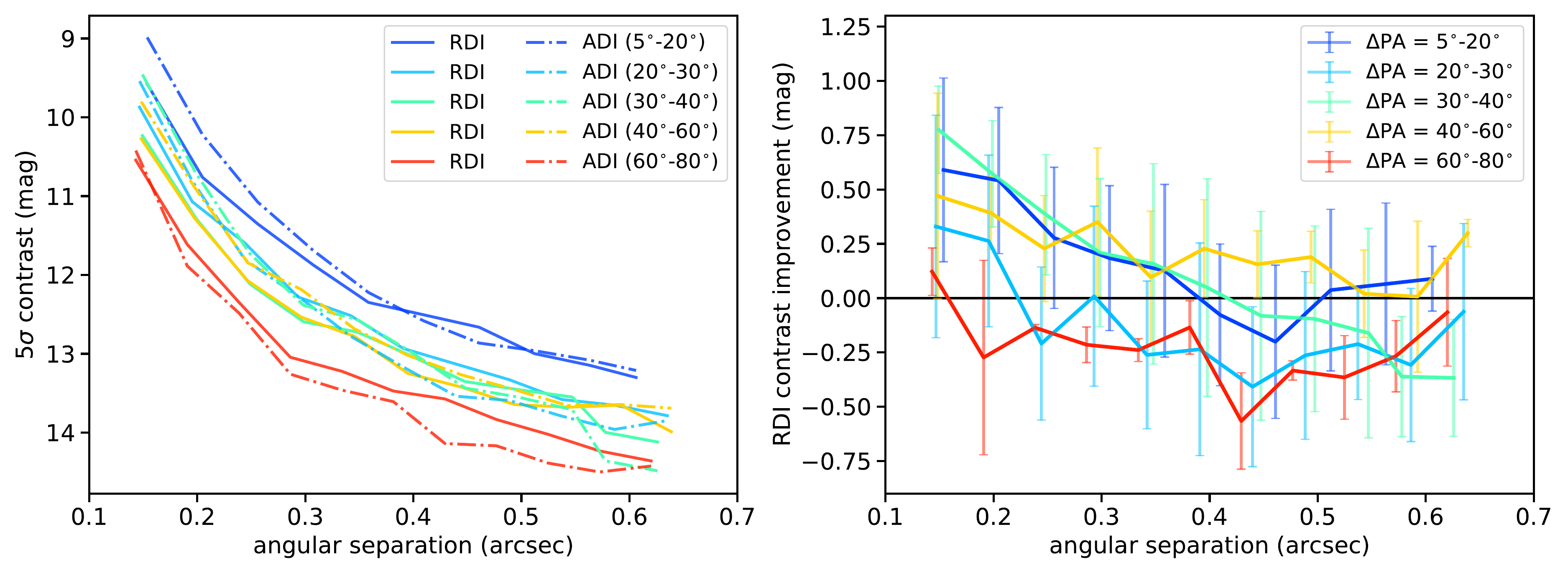}
      \caption{Comparison of RDI and ADI contrasts. \textsl{Left panel:} Contrast curves (5$\sigma$) for the selected sample with the RDI (solid lines) or ADI (dashed lines) reduction, averaged over targets in each bin of PA rotations. \textsl{Right panel:} RDI contrast improvement over ADI. The error bar represents the scatter of the RDI improvements among targets in each bin. Five bins of PA rotations were used.
              }
         \label{Fig:RDI_vs_ADI_delta_PA}
  \end{figure*}
  
   \begin{figure*}
  \centering
    \includegraphics[width=0.98\textwidth]{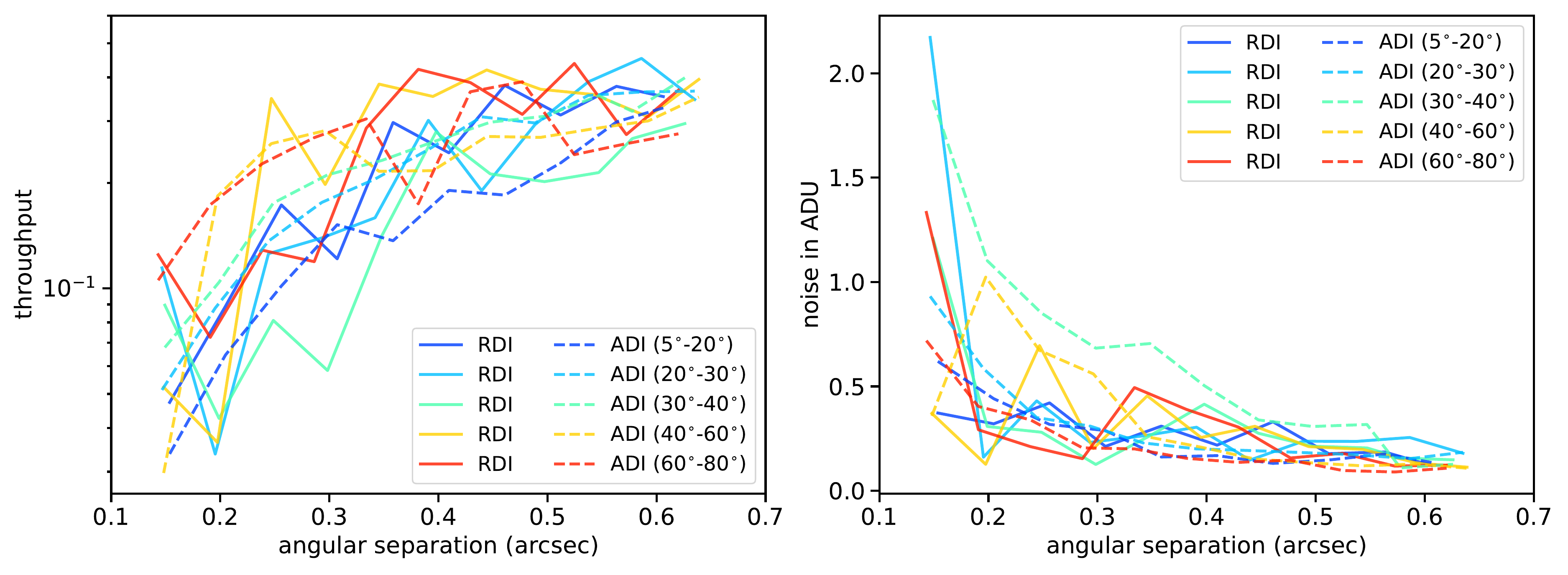}
      \caption{The throughput and noise correspond to the contrast curve shown in Fig.\ref{Fig:RDI_vs_ADI_delta_PA}. \textsl{Left panel:} Throughput of the PCA subtraction using RDI (solid lines) or ADI (dashed lines) as a function of angular separations. \textsl{Right panel:} Noise in the residual image processed with RDI (solid lines) or ADI (dashed lines) as a function of angular separations. Five bins of PA rotations were used. The throughput and noise were averaged over targets in each bin of PA rotations.
              }
         \label{Fig:RDI_vs_ADI_delta_PA_throughput_noise}
  \end{figure*}

The previous analysis was performed on targets with PA rotations of $28.9^{\circ}~\pm~5.8^{\circ}$. However, the PA rotation has a strong impact on the ADI performance. To illustrate the impacts of PA rotations, we selected 20 targets out of 725 observations in our master reference library based on the criteria described in Sect.~\ref{subsec:sample_selection} and listed in Table~\ref{table:sample_delta_PA}. Each observation has similar observing conditions (i.e., seeing, wind speed, Strehl ratio, $H$-band magnitude, and AO loop frequency) that are about the median conditions shown in Fig.~\ref{Fig:distribution_ob_condition}. Only the PA rotations are significantly different and in a range of 5$^{\circ}$ to 80$^{\circ}$, which is divided into five bins. 
As in Sect.~\ref{subsec:RDI_vs_seeing}, we estimated the contrast curves for the selected targets and took the mean of the contrast curves of the test targets in each bin of PA rotation, shown in Fig.~\ref{Fig:RDI_vs_ADI_delta_PA}. The corresponding improvement of RDI over ADI as a function of PA rotations is also shown in Fig.~\ref{Fig:RDI_vs_ADI_delta_PA}. The corresponding uncertainty was estimated by taking the standard deviation of results for targets in each bin at the given separation. 

As shown in Fig~\ref{Fig:RDI_vs_ADI_delta_PA}, both RDI and ADI show that deeper contrasts are achieved for observations under large PA rotations. Observations with large PA rotations usually have a relatively long exposure time. Furthermore, the combination of frames with large PA rotations whitens the noise \citep[i.e., removes the correlated component of the speckle and causes the residual noise to become closer to a Gaussian distribution;][]{Marois2008, Mawet2014} and thus improves the contrast. Similar to the results shown in Sect.~\ref{subsec:RDI_vs_seeing}, RDI can outperform ADI at small angular separations (i.e., \textless0.4\arcsec) for observations with PA rotations of 5$^{\circ}$-60$^{\circ}$. \cite{Wahhaj2021} also reported that RDI outperforms ADI at separations \textless0.4\arcsec~based on the SPHERE star-hopping data. At a separation of 0.15\arcsec, RDI reaches a peak gain of $0.8 \pm 0.2$ mag over ADI for PA rotations between 30$^{\circ}$-40$^{\circ}$. 
RDI shows no or limited gain over ADI for PA rotations of 60$^{\circ}$-80$^{\circ}$, which is expected as the self-subtraction effect can be mitigated by large PA rotations. However, the largest RDI gain over ADI is achieved for PA rotations of 30$^{\circ}$-40$^{\circ}$, not 5$^{\circ}$-20$^{\circ}$\footnote{The median PA rotation is 14.3$^{\circ}$ for targets in a PA bin of 5$^{\circ}$-20$^{\circ}$.}. Their difference is within the scatter of the sample, however.

Fig.~\ref{Fig:RDI_vs_ADI_delta_PA_throughput_noise} shows the throughput of the PCA subtraction and noise in the residual image after the processes of RDI and ADI, which correspond to the contrast curves shown in Fig.~\ref{Fig:RDI_vs_ADI_delta_PA}. Both RDI and ADI throughputs increase from about 5\% to 45\% as increasing angular separations for all bins of PA rotations, which are very similar to observations with different seeing conditions in Sect.~\ref{subsec:RDI_vs_seeing}. For most of the angular separations, the throughput of ADI tends to be higher if an observation has a larger PA rotation, which is a direct demonstration of the self-subtraction effect. Moreover, no significant trend between the RDI throughput and the PA rotation can be found because RDI does not have the self-subtraction effect. The throughput and noise in RDI are coupled, as shown in Eq.~\ref{equ:contrast}, and are only affected by oversubtraction. As expected, noise in the RDI and ADI images decreases with the increase in angular separations because of the properties of speckle noise and photon noise.

\subsection{Impact of the size of the master reference library}
\label{subsect:impact_of_m_size}

  \begin{figure}
  \centering
    \includegraphics[width=0.48\textwidth]{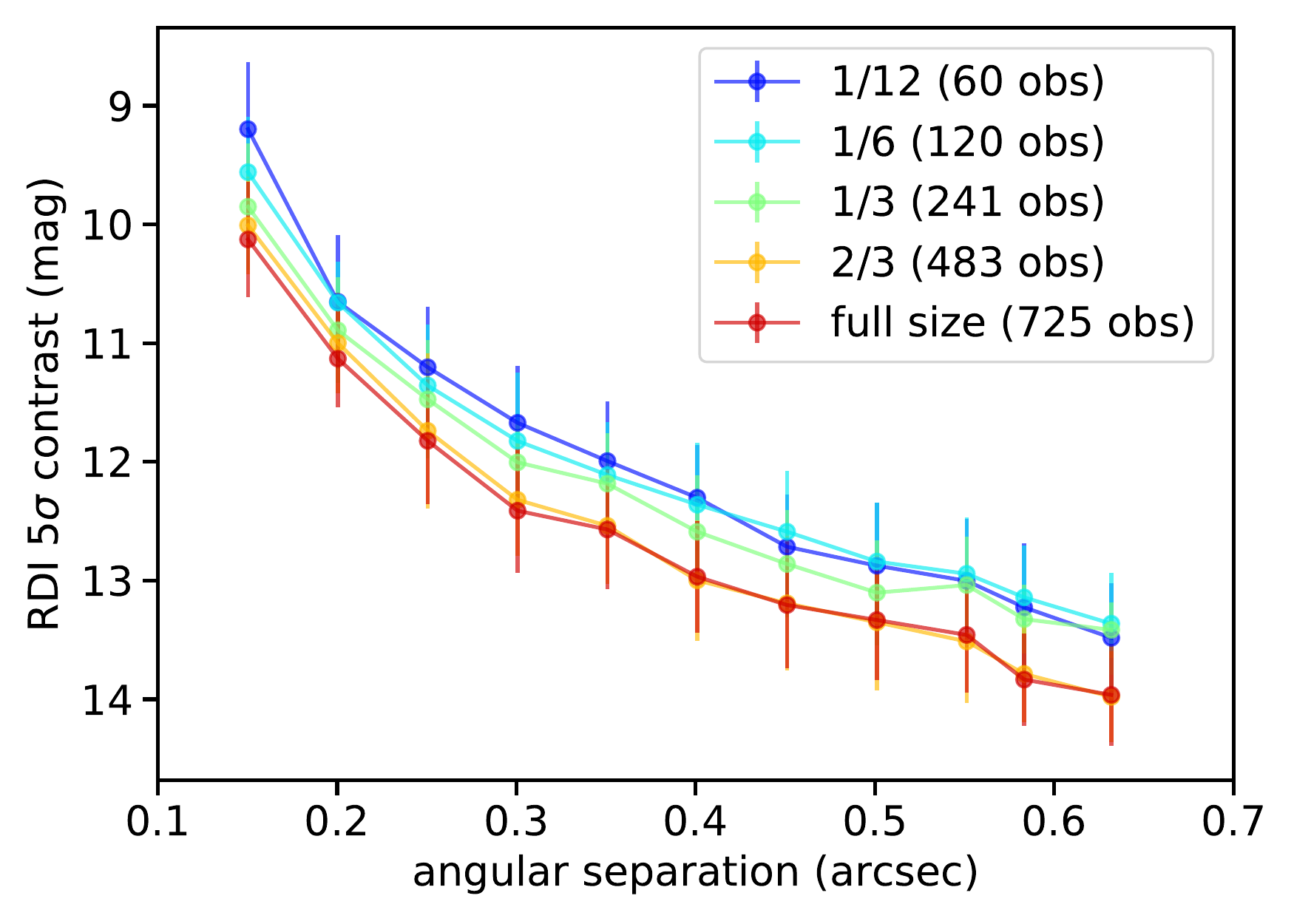}
      \caption{Contrast curves in 5$\sigma$ obtained with RDI using different sizes of master reference library. The fraction and number of observations contained in the master reference library is shown in the figure. A gain of 1 mag at 0.15\arcsec\,separation can be achieved by increasing the size of the master reference library from 60 observations (1/12) to 725 observations (full size).
              }
         \label{Fig:RDI_vs_m_size}
  \end{figure}

To optimize the subtraction of stellar contributions, the master reference library should contain enough coronagraphic images to match given science images obtained under all conditions.  
Space-based instruments (e.g., \textsl{HST}/STIS or \textsl{HST}/NICMOS) usually have stable PSFs and require only thousands of images in the master reference library \citep{ALICE_Choquet2014SPIE}. Conversely, for ground-based instruments, the master reference library is expected to be much larger than that of space-based instruments. This is because the rapid atmospheric turbulence makes RDI less effective in providing representative stellar coronagraphic images. As demonstrated in Fig.~\ref{Fig:RDI_vs_ADI_delta_Seeing}, RDI can outperform ADI for observations at about median conditions of the master reference library. This indicates that more well-matching reference images can improve the PCA subtraction and lead to a deeper contrast. However, the impact of the size of the master reference library remains unclear. In other words, designers of an RDI observation or survey need to know whether they need more reference images to further improve their detection limit. The large size of our master reference library (725 observations with $\sim$$7\times10^{4}$ frames per band) enables us to investigate the impact of the size of the master reference library by dividing our library into subsets. 

To estimate the RDI performance with a smaller master reference library, we randomly extracted 1/12, 1/6, 1/3, and 2/3 of observations from our master reference library, yielding four subsets that contained 60, 120, 241, and 483 observations. The corresponding numbers of reference images per band are about $9\times10^{3}$,  $1.2\times10^{4}$, $2.6\times10^{4}$, and $4.4\times10^{4}$, respectively. A medium-sized RDI survey may contain about 60 reference targets (e.g., \citealp[SHARDDS;][]{Milli2017_SHARDDS} and \citealp[YSES;][]{YSES_Bohn2020MNRAS}). 
We adopted the selected sample from Sect.~\ref{subsec:RDI_vs_seeing}, but only focused on seeing conditions within 0.4\arcsec-0.6\arcsec, yielding a subsample of 8 targets in this test (see also Table~\ref{table:sample_delta_Seeing}). Then we estimated the optimal contrast curves for each target as described in Sect. \ref{subsec:contrast_curve}. The only difference was the size of the master reference library. For the down-selected reference library, we adopted sizes ranging from 200 to 10000 (if possible) images, as described in Sect.~\ref{subsec:contrast_curve}. The final contrast curve of RDI for a given size of a master reference library was formed by averaging over the contrasts of the selected 8 targets. The corresponding uncertainty at each separation was estimated by taking the standard deviation of the contrasts from the selected 8 targets.

Fig.~\ref{Fig:RDI_vs_m_size} shows the RDI performance with the different sizes of the master reference library. We achieved a gain of $\sim$1 mag at 0.15\arcsec\,separation and gains of 0.5-0.8 mag between separations of 0.20\arcsec-0.65\arcsec \ when we increased the size of the mater reference library from 1/12 to full size. This indicates that the performance of an RDI survey with 60 reference targets can be further improved by about 1 mag when we include all the archival data. Although the uncertainty is about 0.5 mag, the RDI performance has systematic gains across all separations (0.15\arcsec-0.65\arcsec) with increasing size of our mater reference library from only 1/12 to full size. 
A larger master reference library statistically provides more well-matching frames, which in turn improves the accuracy of the PCA subtraction. Despite the fluctuation of gains at larger separations (\textgreater0.35\arcsec), RDI shows a systematic gain at short separations (\textless0.35\arcsec). This indicates that a better RDI performance at short separations can be expected with the accumulation of SPHERE data in the future. Overall, Fig.~\ref{Fig:RDI_vs_m_size} proves the necessity of a large number of references in the archive to obtain good RDI performance.

\subsection{RDI performance as a function of the down-selection of the reference library}
\label{subsect:RDI_vs_size}

  \begin{figure*}
  \centering
    \includegraphics[width=0.98\textwidth]{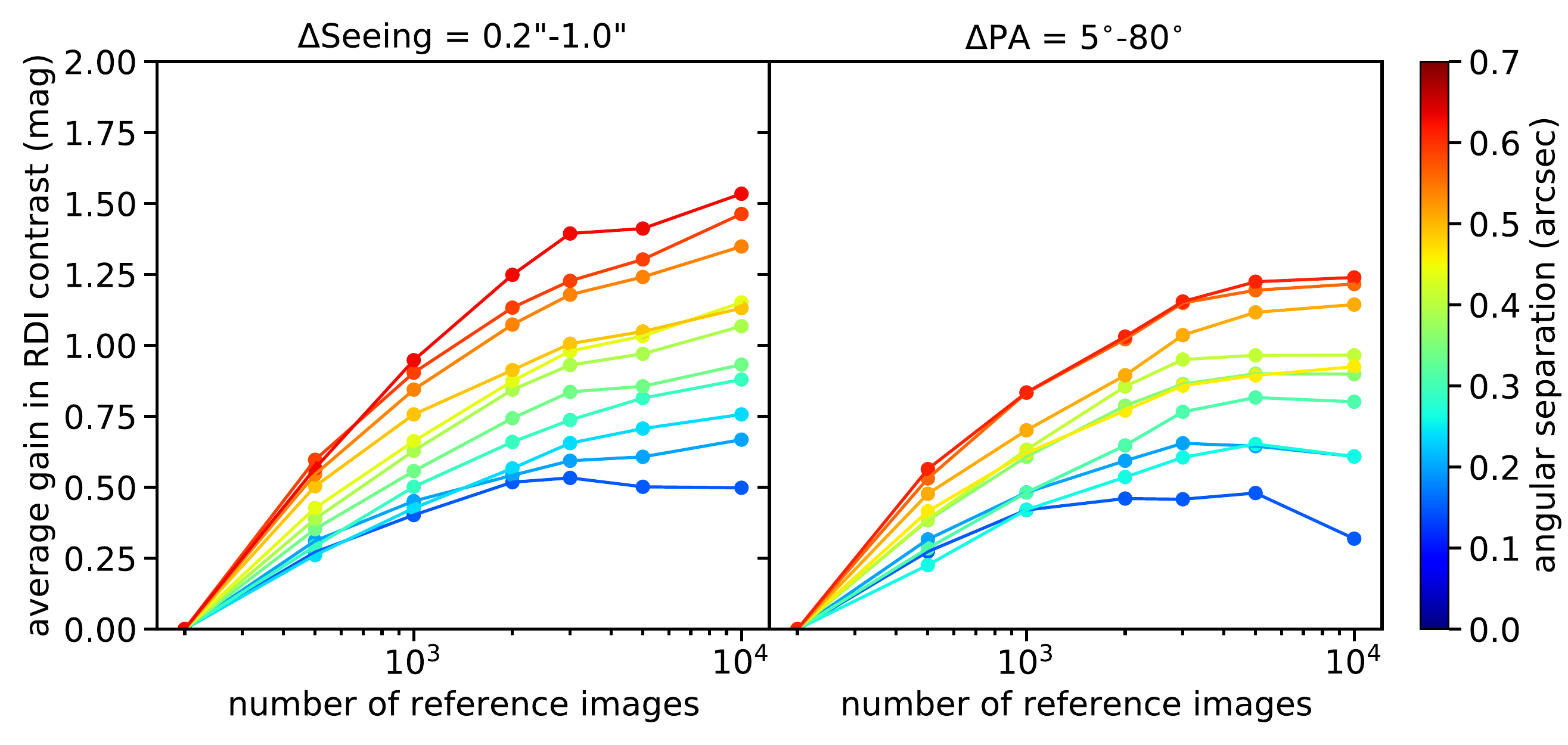}
      \caption{Average gain in RDI 5$\sigma$ contrast as a function of reference library size used to subtract stellar contributions. RDI contrasts were averaged over all the targets with similar observing conditions, but different seeing conditions (\textsl{left}) or different PA rotations (\textsl{right}). For all the angular separations, RDI contrasts improved with increasing number of reference images and reached plateaus after a library size of $\sim$3000 reference images ($\sim$4.5\% of all the references). 
              }
         \label{Fig:RDI_average_gain_ref_size}
  \end{figure*}

The size of the reference library can also affect the RDI performance. In our RDI reduction, we down-selected references from the master reference library to form the reference library. By selecting the most correlated references to each science image, we  optimized the PCA subtraction of the stellar contributions. Unlike Sect.~\ref{subsect:impact_of_m_size}, here we used the full size of the master reference library for the down-selection. 
Using two target samples from Sects.~\ref{subsec:RDI_vs_seeing} and \ref{subsec:RDI_vs_PA}, we investigated the average performance of RDI for a given size of the reference library.  The adopted size of the reference library ranges from very small and selective (200 images, i.e., 0.3\% of all the references) to large and more diverse (10000 images, i.e., 14.5\% of all the references).  For each observation at each angular separation, we compared the improvement of RDI contrast with respect to the RDI contrast obtained with a library size of 200 images. Then we averaged over all the targets in each sample to obtain the average RDI gain for a given angular separation.

Fig.~\ref{Fig:RDI_average_gain_ref_size} shows the average RDI gain in contrast as a function of reference library size, derived from targets with different seeing conditions and different PA rotations listed in  Tables~\ref{table:sample_delta_Seeing} and \ref{table:sample_delta_PA}. The performance of RDI systematically improves with the size of libraries and reaches the plateau of contrast gain where the library size is larger than about 3000-5000 images (4.3\% --7.2\% of all the references). This suggests that including more good references can indeed optimize the PCA subtraction. This is due to the nature of the reconstruction of the stellar model using PCA. An infinite number of references is needed to reconstruct the given science image exactly \citep{Soummer2012}. However, without additional observations, adding more references can only include references that are less correlated with a given science image. Therefore, we reach a plateau of contrast gain after using roughly 5\% of most correlated images from the master reference library. As a comparison, \textsl{HST}/NICMOS achieved its optimal performance with 30\% to 80\% of most correlated images from its master reference library \citep{ALICE_Choquet2014SPIE}.

Furthermore, increasing the size of the reference library provides higher gains at large separations (up to 1.5 mag at 0.65\arcsec) than at short separations (about 0.5 mag at 0.15\arcsec), as shown in Fig.~\ref{Fig:RDI_average_gain_ref_size}. Less improvement at short angular separations may be caused by the large variation of speckles, which cannot be completely removed by RDI using archival data without oversubtraction. Observing references and science targets nearly simultaneously can help to remove these speckles and obtain better RDI performance at short angular separations, as demonstrated by the star-hopping mode on SPHERE \citep{Wahhaj2021}. In contrast, the speckles at large separations are relatively stable and hence can be more easily modeled and subtracted when enough good references are available. 
Interestingly, both samples show similar behaviors. This indicates that the RDI gains as a function of reference library size that we observed do not depend on seeing conditions or PA rotations.

\section{Circumstellar disk imaging in total intensity with RDI}
\label{sec:disk_imaging}

  \begin{figure}
  \centering
    \includegraphics[width=0.46\textwidth]{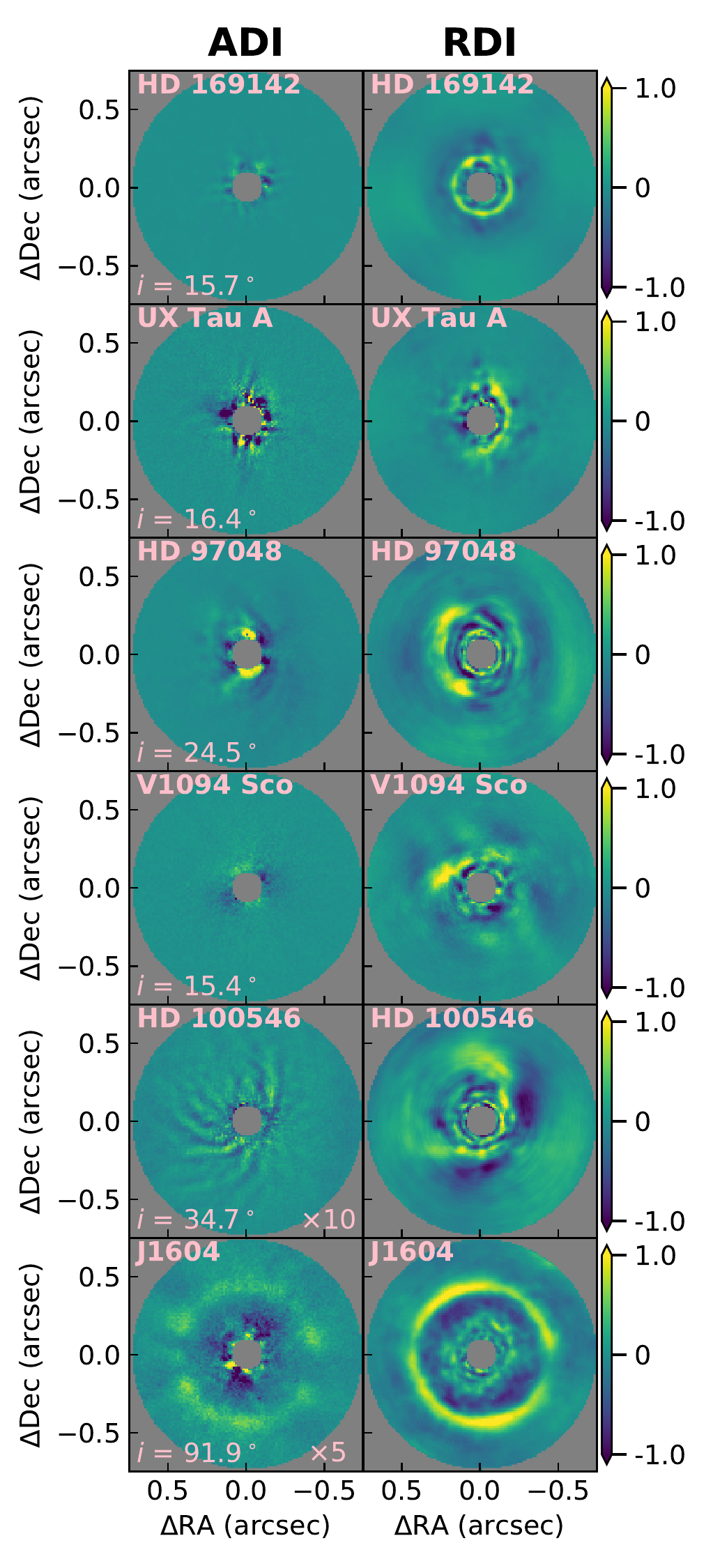}
      \caption{Comparison of SPHERE/IRDIS disk images in $H23$ reduced with ADI (\textsl{left}) and RDI (\textsl{right}). An arbitrary linear color scale is applied, which is normalized to the maximum flux in the RDI image of each target. Because the ADI images of HD\,100546 and J1604 severely suffer from flux loss due to self-subtraction, we linearly scaled these ADI images by factors of 10 and 5, respectively. The star is located in the center of each image.  The central gray region shows the position of the coronagraph mask with a diameter of 196 mas (16 pixels). The adopted FoV is 0.735\arcsec~in radius.
              }
           \label{Fig:disk_ADI_vs_RDI}
  \end{figure}

    \begin{figure*}
  \centering
    \includegraphics[width=0.91\textwidth]{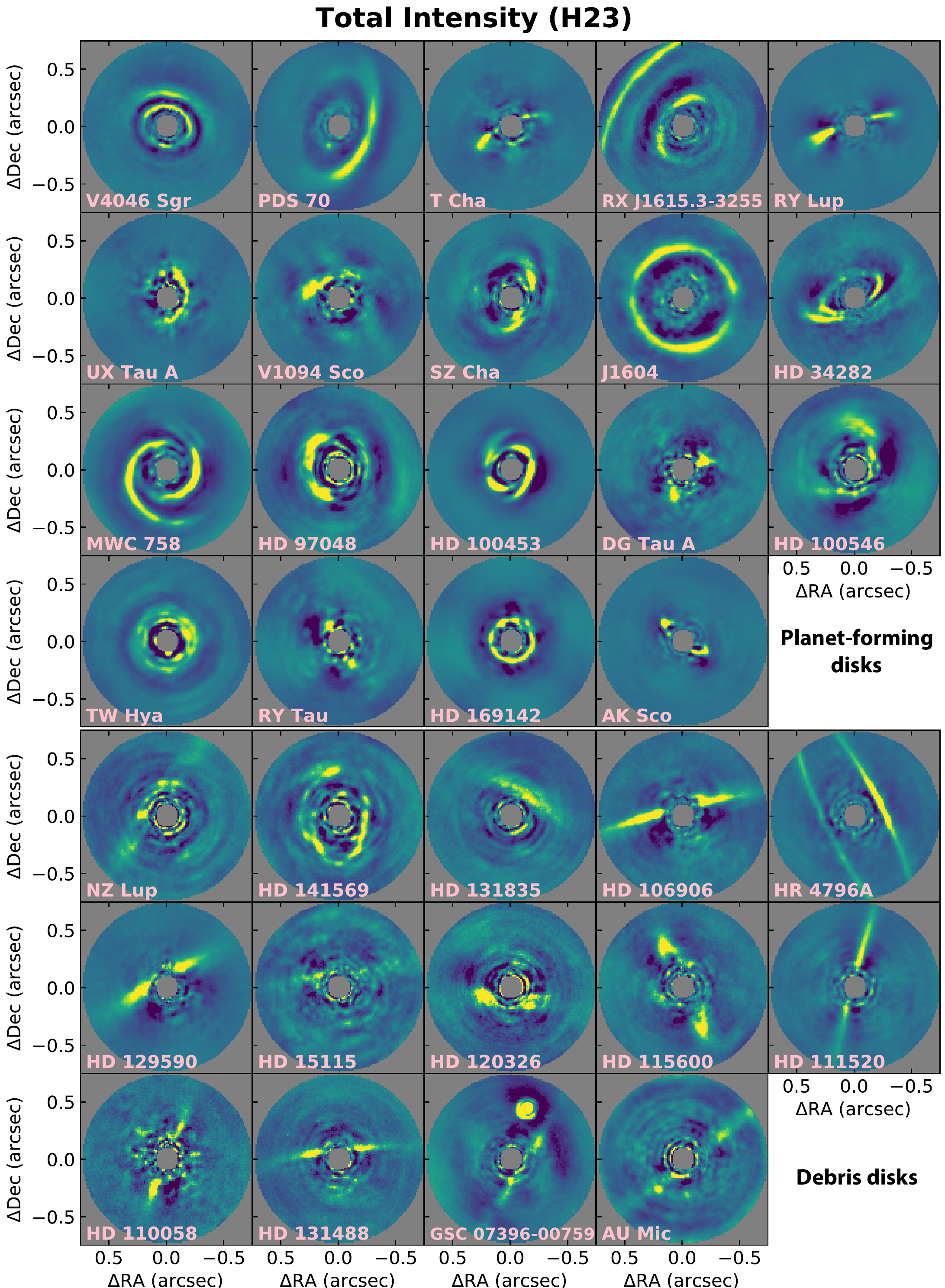}
      \caption{SPHERE/IRDIS disk images in total intensity after the process of RDI. All panels are displayed in linear color scales. Due to different disk surface brightnesses, we linearly scaled some disk brightnesses to improve the visibility of the substructure. The star is located in the center of each image. The central gray region shows the position of the coronagraph mask with a diameter of 196 mas (16 pixels). The adopted FoV is 0.735\arcsec~in radius. For consistency, we did not increase our FoV for disks extending outside of the FoV. We note that RDI does not have a FoV limit.
      }
           \label{Fig:RDI_disks}
  \end{figure*}

\label{sect:disk_imaging}

The RDI technique can not only outperform the ADI technique at short separations for point-source detection but also has advantages for disk imaging. Currently, ADI is adopted as the primary technique for most observations and surveys of disks in total intensity. It is well known, however, that ADI is limited by the self-subtraction effect that may lead to poor recoveries of disk morphology and surface brightness \citep{Milli2012}. The ALICE project \citep{Soummer2014, ALICE_Choquet2014SPIE} demonstrated that applying RDI on space-based instruments can provide a gain in disk imaging. In this section, we demonstrate the capability of SPHERE in the detection of circumstellar disks by using RDI and compare it with that of ADI.

\subsection{Robust recovery of disk features with RDI}

To illustrate the performance of RDI in disk imaging, we selected targets with known circumstellar disks that were not or poorly detected by ADI, as shown in Fig.~\ref{Fig:disk_ADI_vs_RDI}. %
The ADI images of selected six targets\footnote{For more details about the targets, we refer to the following papers: \object{HD\,169142}: \cite{Gratton2019}; \object{UX\,Tau}: \cite{Menard2020}; \object{HD\,97048}: \cite{Ginski2016}; \object{HD\,100546}: \cite{Rameau2017}; \object{V1094\,Sco}: \cite{van_Terwisga2018}; and \object{2MASS\,J16042165-2130284} (hereafter J1604): \cite{Pinilla2018}.} show no or limited disk features due to the self-subtraction effect, while RDI images reveal more features. To avoid aggressive post-processing in ADI, we only used 5\% of the number of temporal frames as the number of subtracted PCs for UX\,Tau\,A and V1094\,Sco. For the rest of the targets in Fig.~\ref{Fig:disk_ADI_vs_RDI}, we adopted a fraction of 2\% for subtracted PCs in ADI reductions. For RDI, we used 20\% of the number of reference images as the number of PCs in RDI for all the targets in Fig.~\ref{Fig:disk_ADI_vs_RDI}. The parameters used in the RDI reduction are listed in Table~\ref{table:disk_list}.

The ADI image is sensitive to the number of subtracted PCs due to the self-subtraction effect. Even when only 2\% of the PCs are subtracted, the self-subtraction effect still strongly affects the surface brightness of disks detected by ADI, reducing the disk emission (see HD\,100546 and 2MASS\,J16042165-2130284 in Fig.~\ref{Fig:disk_ADI_vs_RDI}). For example, all the disk signals of 2MASS\,J16042165-2130284 (hereafter J1604) become too faint to be considered as a detection when the subtracted PCs are larger than 10\% using ADI. The disk signal of HD\,100546 also has similar behavior that decreases very fast when the subtracted PCs change from 2\% to 10\%. The remaining images show no significant disk signal when no more than 5\% of the PCs are subtracted, probably due to the self-subtraction effect caused by small PA rotations (\textless25$^{\circ}$).

In contrast, the disks in RDI images remain visible with consistent features as the PCs increase from 2\% to 50\%, providing more robust recoveries of disk features. \cite{Ruane2019} analyzed the RDI performance on Keck/NIRC2 data and found consistent results: the ADI image was sensitive to the number of subtracted PCs, while the RDI image was not. The insensitivity to subtracted PCs can help avoid the bias in post-processing that is caused by the fine-tuning of the parameters. This increases the robustness of the RDI detection.

The comparisons between ADI and RDI in Fig.~\ref{Fig:disk_ADI_vs_RDI} show the improvement in disk imaging after the removal of the self-subtraction effect. In addition to reducing the surface brightness of disks, the self-subtraction effect can also sharpen the azimuthal features \citep{Milli2012}. For example, the disk signal of J1604 is self-subtracted after ADI, which alters the real disk features such as disk shadowing \citep{Pinilla2018} by introducing artifacts (e.g., more dips). For HD\,100546 in Fig.~\ref{Fig:disk_ADI_vs_RDI}, the disk features revealed in our ADI image are similar to the ADI image in \cite{Garufi2016} and \cite{Sissa2018} that used the same data set. Our RDI image is similar to the GPI $H$ band images that were reduced with RDI by  \cite{Rameau2017}. Assessing the confidence of recovered disk features after the PCA subtraction requires disk modeling, which is beyond the scope of this paper. The difference between ADI and RDI images shows one of the main limitations in ADI in precisely recovering geometric parameters and photometry. One of the solutions might be applying disk forward modeling in ADI \citep{Pueyo2016, DiskFM_Mazoyer2020SPIE}. Alternatively, RDI could be a natural solution because it does not have the self-subtraction effect.

Although both ADI and RDI suffer oversubtraction caused by the PCA subtraction, most of the disk signal is self-subtracted in ADI with no more than 5\% of PCs.  Moreover, oversubtraction is limited in RDI images even at short angular separations, as shown in Fig.~\ref{Fig:disk_ADI_vs_RDI}. It is worth noting that the disk features are clearly visible even when 50\% of PCs are subtracted. This mild oversubtraction effect in PCA can be corrected for by forward modeling \citep{Pueyo2016} to accurately characterize the surface brightness. Providing the throughput estimation is very difficult for disk imaging due to the complex disk morphology and requires detailed disk modelings. No throughput correction was made in the disk images presented in this work. We will cover the disk imaging with detailed disk modeling and analysis in future works. In summary, applying RDI on SPHERE/IRDIS data can provide a more robust recovery of disk features in total intensity than ADI.

\subsection{Higher disk detection rate with RDI than ADI}

After processing archival SPHERE/IRDIS data in the $H23$ band with ADI and RDI, we detected 33 circumstellar disks in total intensity, including 19 planet-forming disks and 14 debris disks. The RDI detection of the disks in this work is shown in Fig.~\ref{Fig:RDI_disks} and summarized in Table~\ref{table:disk_list}. The size of the reference library and the number of PCs subtracted in the RDI reduction are also listed in Table~\ref{table:disk_list}.
The circumstellar disks of \object{DG\,Tau\,A} and HD\,131488 are resolved in scattered light for the first time. The detailed analysis of the disk in HD\,131488 will be presented in Pawellek~et~al.~(in~prep.). To our best knowledge, three disks (\object{V1094\,Sco}, \object{UX\,Tau\,A}, and \object{SZ\,Cha}) are detected in total intensity for the first time. The detailed analysis of the disk SZ\,Cha will be presented in Hagelberg~et~al.~(in~prep.). 

In addition to new detections, RDI recovers more disk features than previous ADI results. We detect the inner ring~1 at $\sim$0.25\arcsec~around HD\,97048 in total intensity, which was not or only marginally visible in the ADI image \citep{Ginski2016}. In HD\,141569, we detect more disk emission in the west side of the ring R3 than the ADI detection in \cite{Singh2021}. We also detect a large fraction of the full ring including the backside for AK Sco, while the previously published ADI result \citep{Janson2016} was only able to see the forward-scattered side. The detailed analysis of the disks around HD\,110058, HD\,111520, and HD\,120326 will be presented in Stasevic~et~al.~(in~prep.) and Desgrange~et~al.~(in~prep.).

In addition to recovering more disk features, RDI also has a higher detection rate of disks than ADI. Of these 33 disks, 4 disks (\object{TW\,Hya}, V1094\,Sco, UX\,Tau\,A, and HD\,169142) are only detected in the RDI reductions. All the disks with a nondetection in ADI have low inclinations (\textless50$^{\circ}$), which may lead to severe self-subtraction. The debris disk survey from GPIES also found that an increasing disk inclination can improve the capability of ADI to detect disks in total intensity \citep{Esposito2020}. Our work shows that RDI potentially has a higher detection rate than ADI in disk imaging, especially for low-inclination disks. We note that our field of view is 0.735\arcsec~in radius. It is possible that disk features can be detected outside our field of view and are not listed in Table~\ref{table:disk_list}.

\section{Discussions and conclusions}
\label{sec:conclusions}

We have presented a thorough analysis of the RDI performance on SPHERE/IRDIS in the direct imaging of exoplanets and circumstellar disks. We made use of all public archival data observed in the past 5 years by IRDIS in the $H23$ band with an unprecedented size of reference images to perform RDI, including about $1.4 \times 10^{5}$ reference images from 725 observations. The results of averaged RDI performance on SPHERE/IRDIS are summarized below.

\begin{enumerate}

\item Reference-star differential imaging can outperform ADI at small angular separation (\textless0.4\arcsec) if the observing conditions are approximately the median conditions of our master reference library (i.e., seeing in 0.4\arcsec-0.6\arcsec). At a separation of 0.15\arcsec, RDI outperforms ADI by $0.8 \pm 0.3$ mag on average for observations under median conditions. Furthermore, our target sample shows that the RDI performance is better than or equal to that of ADI within separations of 0.4\arcsec~for observations with PA rotations lower than 60$^{\circ}$.

\item In the point source detection, we demonstrated that including more observations (i.e., more references) in the master reference library indeed helps to improve the performance of RDI at separations of 0.15\arcsec-0.65\arcsec. An average gain of $\sim$1 mag can be achieved at 0.15\arcsec\,separation by increasing the number of observations in the master reference library from 60 (typical survey size) to 725 (full archive). 

\item In the point source detection, we find that increasing the number of reference images to reconstruct the stellar image with PCA help to improve the average performance of RDI for all tested angular separations. The average performance of RDI shows no or limited improvement when more than 3000 reference images are used. This suggests that a typical library size for obtaining an optimal RDI reduction is about 3000 - 5000 images for a search for point-like sources.

\item In disk imaging, RDI does not have the self-subtraction effect and reveals more features than ADI. The disk features in RDI images are insensitive to subtracted PCs. Hence, RDI provides a more robust recovery of the disk morphology.

\item We systematically processed and presented 33 circumstellar disks in total intensity obtained by SPHERE/IRDIS in $H23$. The circumstellar disks of DG\,Tau\,A and HD~131488 are resolved in scattered light for the first time. To our best knowledge, three disks (V1094\,Sco, UX\,Tau\,A, and SZ\,Cha) are detected in total intensity for the first time.

\item Reference-star differential imaging has a better capability of detecting disks than ADI, especially for low-inclination (\textless 50$^{\circ}$) disks. Four of the 33 disks detected in this work are only detected in RDI images, not in ADI images.

\end{enumerate}

Our successful application of RDI on SPHERE data shows that RDI can be applied to ground-based surveys of exoplanets with better performances than ADI at short separations. More importantly, our RDI strategy does not require additional observation of calibration or reference stars. Therefore, it can easily be adopted into legacy or future SPHERE surveys. Furthermore, for a medium-size RDI survey with 60 reference targets, implementing our master reference library can further improve the RDI performance by $\sim$1 mag at 0.15\arcsec\,separation, as demonstrated in Sect.~\ref{subsect:impact_of_m_size}.

We showed that the self-subtraction effect on disks that were previously processed with ADI can be overcome in the disk imaging. No additional observations are needed. The self-subtraction effect limits the detection of low-inclination disks in total intensity \citep{Esposito2020} and sharpens azimuthal features. However, RDI does not have such issues, as shown in Fig.\ref{Fig:disk_ADI_vs_RDI}, which is expected. Future observations or surveys can adopt our master reference library and perform RDI in their disk observations to obtain disk images in total intensity. 

  \begin{figure}
  \centering
    \includegraphics[width=0.48\textwidth]{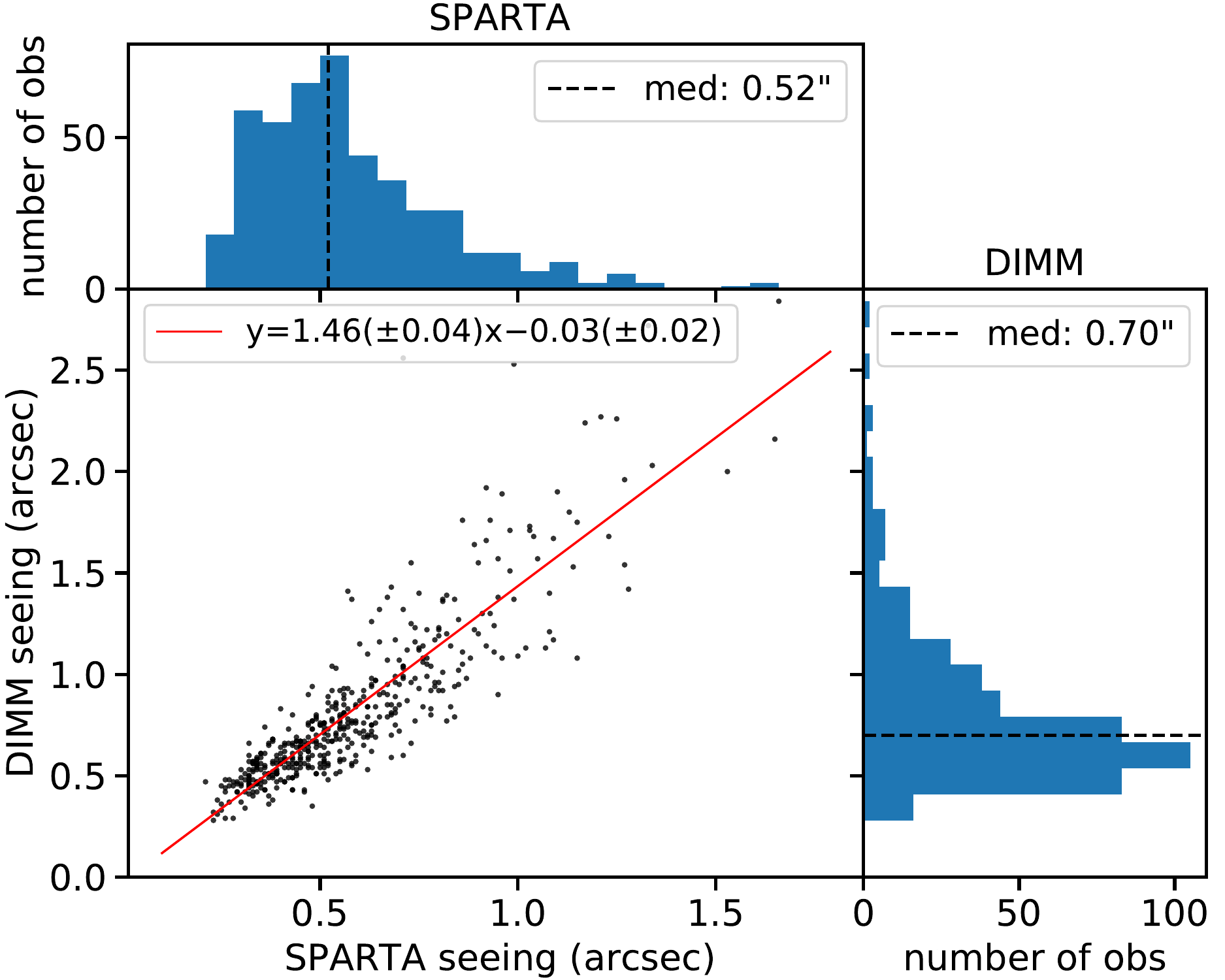}
      \caption{Comparison of the SPARTA seeing and the DIMM seeing for observations in our master reference library. We only adopted observations after the upgrade of the Paranal ASM system in April 2016. The median DIMM seeing for our reference library is 0.7\arcsec. }
           \label{Fig:seeing_vs_seeing}
  \end{figure}

In general, we recommend future observations be carried out under median observing conditions, as shown in Fig.~\ref{Fig:distribution_ob_condition}, to achieve the optimal RDI gain over ADI. However, in practice, observers can only specify the required seeing conditions\footnote{SPHERE defined six turbulence categories for users to choose from, which consist of a few combinations of seeing and coherence time.} to meet their observational goal. 
Since the upgrade in April 2016, the Paranal Astronomical Site Monitoring (ASM) has the new MASS-DIMM system \citep{Sarazin1990, Kornilov2007}, which  provides the seeing conditions at zenith for SPHERE to decide whether an observation can be executed. In Fig.~\ref{Fig:seeing_vs_seeing} we compare the line-of-sight seeing measured by the SPHERE real-time computer system (SPARTA) and the seeing measured by the MASS-DIMM system. 
Although the two systems measured the turbulence at different spatial locations with different ground effects (i.e., with and without a dome), the correlation is still tight, shown in \cite{Milli2017_AO4ELT5}. The median DIMM seeing for our reference library is 0.7\arcsec. Excellence seeing conditions (i.e., DIMM seeing \textless0.6\arcsec) lead to deeper contrasts for both RDI and ADI, but they are also an expensive request. As a technique alternative to ADI, RDI has optimal gains over ADI for DIMM seeing conditions of 0.6\arcsec-0.8\arcsec (see also Fig.\ref{Fig:RDI_vs_ADI_delta_Seeing}).   
Therefore, we recommend future observations be carried out under seeing conditions of 0.6\arcsec-0.8\arcsec. Such seeing conditions correspond to the turbulence categories of 20\% and 30\%.

In addition, using our reference library to perform RDI requires the same coronagraph settings as for the coronagraph in the $\texttt{N\_ALC\_YJH\_S}$ configuration and the $\texttt{DB\_H23}$ filter pair. We will present the RDI library for the $\texttt{DB\_K12}$ filter pair in future works. By adopting RDI in the survey of SPHERE or other ground-based instruments, we can avoid several limitations introduced by ADI. As a result, the design of a survey or observation can be more flexible, covering a larger declination range without the time constraints caused by the self-subtraction effect. 

We studied the impact of the reference library based on the images processed with PCA. Other methods are available to build and subtract the stellar contribution, such as non-negative matrix factorization \citep[NMF;][]{NMF_Ren2018}. PCA removes the mean of the image. If a bright astrophysical signal (i.e., a bright disk) is present, PCA creates negative regions (i.e., dark features around the disk), which clearly is an artifact. Although such an artifact will not affect the general comparisons between ADI and RDI in Sect.~\ref{sec:disk_imaging}, it may prevent the further application of RDI in characterizing disks. However, this issue can be largely resolved with NMF. 
Furthermore, to avoid overfitting in RDI, \cite{Ren2020_DIsNMF} proposed an improved method, which is data imputation using sequential non-negative matrix factorization (DI-sNMF). In future work, we will use DI-sNMF to demonstrate the performance of our RDI approach in disk imaging (Xie~et~al.~2022,~in~prep.).


SPHERE+ is the proposed upgrade of SPHERE. It is equipped with a pyramid infrared wavefront sensor, an optimized coronagraph, and noncommon path aberrations compensation \citep{SPHERE+_Boccaletti2020}. With a faster AO correction ($\sim$3kHz), SPHERE+ is expected to regularly obtain the same image quality that can currently only be obtained with SPHERE in the best 5\% of observing conditions \citep{SPHERE+_Boccaletti2020}. Better AO correction can provide a more stable PSF, which enhances the performance of RDI. 

The prime aim of SPHERE+ is to image a young Jupiter down to the snow line at $\sim$3 au, bridging the gap with indirect techniques \citep{SPHERE+_Boccaletti2020}. At short angular separations around the IWA of SPHERE+ (\textless100 mas) like this, ADI will be limited by the severe self-subtraction effect. Based on current SPHERE data, we demonstrated that RDI is more sensitive than ADI in searching for planets at short separations (\textless0.4\arcsec). With a more stable PSF and a smaller IWA, it is necessary to consider using RDI as one of the imaging strategies on the future SPHERE+, especially to image the expected peak of gain in the exoplanet population at 3-5~au. 


\begin{acknowledgements}
We thank the anonymous referee for comments that improved the clarity of this work. We thank Dr. David Mary for the beneficial discussion. This research made use of Astropy\footnote{\url{http://www.astropy.org}}, a community-developed core Python package for Astronomy \citep{astropy:2013, astropy:2018}.
The performance analyses are based on observations collected at the European Organisation for Astronomical Research in the Southern Hemisphere under ESO programmes 095.C-0298, 095.C-0346, 095.C-0549, 095.C-0607, 096.C-0241, 097.C-0060, 097.C-0079, 097.C-0826, 097.C-0864, 097.C-0865, 097.C-0949, 097.C-1019, 097.C-1042, 098.C-0739, 099.C-0693, 0100.C-0543, 0101.C-0753, 0104.C-0183, 1100.C-0481, 198.C-0209, and 295.C-5034. We thank all the principal investigators and their collaborators who prepared and performed the observations with SPHERE. Without their efforts, we would not be able to build the master reference library to enable our RDI technique. This publication makes use of data products from the Two Micron All Sky Survey, which is a joint project of the University of Massachusetts and the Infrared Processing and Analysis Center/California Institute of Technology, funded by the National Aeronautics and Space Administration and the National Science Foundation. AV acknowledges funding from the European Research Council (ERC) under the European Union's Horizon 2020 research and innovation programme (grant agreement No.~757561).
\end{acknowledgements}

\bibliographystyle{aa}
\bibliography{ms}

\begin{appendix}

\section{Pointing stability of SPHERE/IRDIS}
\label{subsec:pointing_stability}
Image alignment is one of the key steps in RDI. The ALICE program used the diffraction pattern of the telescope struts to perform the alignment \citep{ALICE_Choquet2014SPIE}. GPI used the positions of satellite spots to align all of the images \citep{Gerard2016}. SPHERE can also generate satellite spots on coronagraph images as GPI by introducing a 2D periodic modulation on the high-order deformable mirror \citep{Beuzit2019}. However, SPHERE usually turns on the satellite spots only at the beginning and/or the end of the science observation to obtain so-called star center images. There is no additional calibration to obtain the star center behind the coronagraphic mask during the science observations and observers usually rely on the pointing stability of SPHERE. 
Even if the instrument were very stable for a given observation, we still need to align such a data set with the master reference library. This was done with our image alignment approach, in which we aligned all the data to a common template image.

We examined the pointing stability of SPHERE/IRDIS during the science observations based on the position offsets we measured in our image alignment process. In this analysis, we used all the data in the master reference library, but excluded the observations with failed star center images. We therefore only focused on completed observations, which yielded a total number of 654 observations. Each temporal frame in a given observation has offsets in the $x$ and $y$ directions with respect to the template we used to align all the images. We took the standard deviations of the offsets in $x$ and $y$ directions to trace the pointing stability of IRDIS in the given observation. 

  \begin{figure}[h!]
  \centering
    \includegraphics[width=0.48\textwidth]{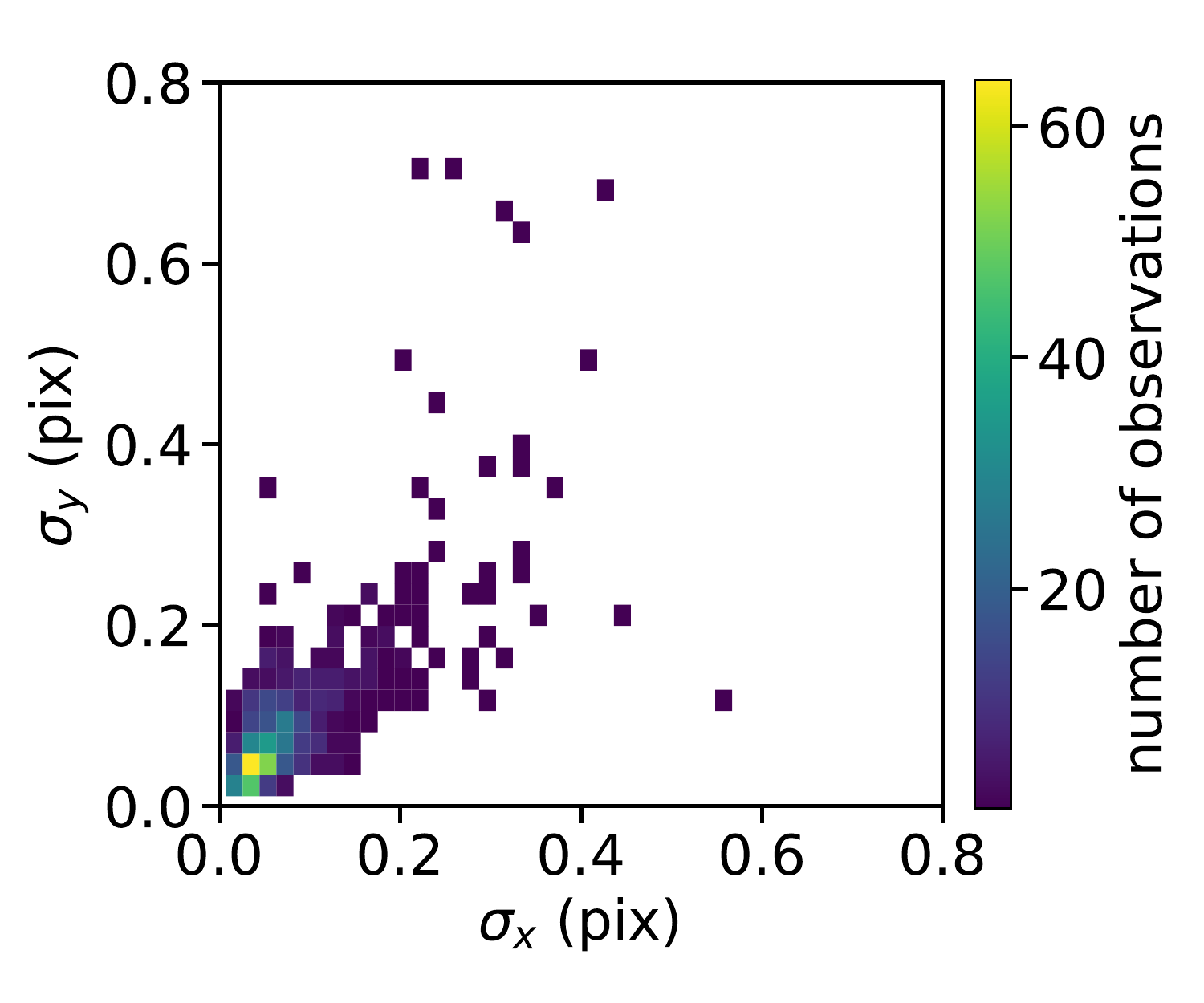}
      \caption{Pointing stability of IRDIS in $H$2 during 654 observations. The median values of the standard deviation of pointing offsets in the $x$ and $y$ directions are about 0.06 pixel and 0.07 pixel, respectively.
              }
           \label{Fig:pointing_stability_IRDIS}
  \end{figure}
\FloatBarrier

Fig.~\ref{Fig:pointing_stability_IRDIS} shows the pointing stability of IRDIS in $H$2 in 654 science observations. The median values of the standard deviation of pointing offsets in the $x$ and $y$ directions are about 0.06 pixel and 0.07 pixel, respectively. Assuming the offsets of a given observation follow a Gaussian distribution, a full width at tenth maximum (FWTM) of 1 pixel requires the standard deviation to be smaller than 0.233 pixel. As shown in Fig.~\ref{Fig:pointing_stability_IRDIS}, over 94\% of observations have both $\sigma_{x}$ and $\sigma_{y}$ smaller than 0.233 pixel, which demonstrates the stability of IRDIS at the subpixel level. Furthermore, the standard deviation of \textless~0.1 pixel corresponds to an FWTM of \textless~0.43 pixel or \textless~5.3 mas on average, assuming the offsets of a given observation follow a Gaussian distribution. For the $H$3 band, we obtain a similar stability, which is expected because $H$2 and $H$3 images were observed simultaneously.

\section{Image alignment}
\label{appendix:frame_registration}
\begin{figure}[h!]
  \centering
    \includegraphics[width=0.48\textwidth]{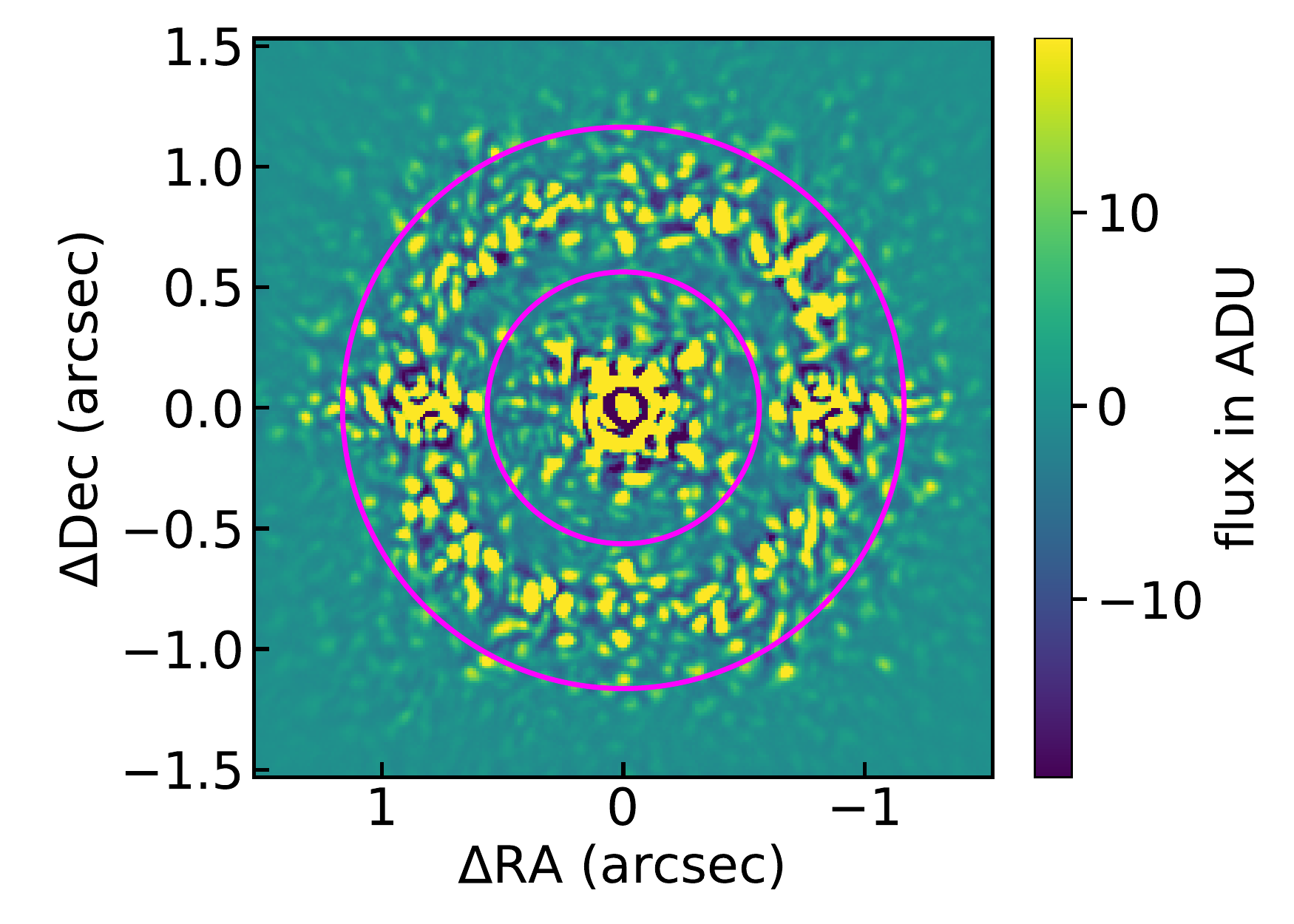}
      \caption{Reference template used in the image alignment. The correction ring is clearly shown in the median-filtered image at around 0.8\arcsec. We only used regions between the two magenta circles in the image alignment.
              }
      \label{Fig:frame_registration}     
  \end{figure}

As mentioned in Sect.~\ref{subsec:building_ref_lib}, we adopted the bright star HD~121156 as our reference template in the image alignment. The median-filtered $H2$ image of  HD~121156 is shown in Fig.~\ref{Fig:frame_registration}. The image mask used in the image alignment is indicated by two magenta circles in Fig.~\ref{Fig:frame_registration}. The image mask was linearly scaled to adopt the larger size of the correction ring in the $H3$ band.

We evaluated the alignment by using Eq.~\ref{equ:MSE}. We compared the reference template with the aligned images that were shifted by offsets provided by Eq.~\ref{equ:loss_function}. The image mask shown in Fig.~\ref{Fig:frame_registration} was used to only focus on the regions of the correction ring in the reference template $S$ and a given image $R$. 
We set a threshold of 90\% of the MSE value of the reference template subtracting an empty image to identify images that failed the image alignment. An empty image indicates an extreme case in which the image was entirely shifted outside the FoV. Such a 90\% threshold was determined by balancing low S/N images and really failed images due to reasons mentioned in Sect.~\ref{subsec:building_ref_lib}.

  \FloatBarrier

\section{Validation of the throughput estimation}
\label{appendix:Validation_throughput_estimation}
\begin{figure}
  \centering
    \includegraphics[width=0.4\textwidth]{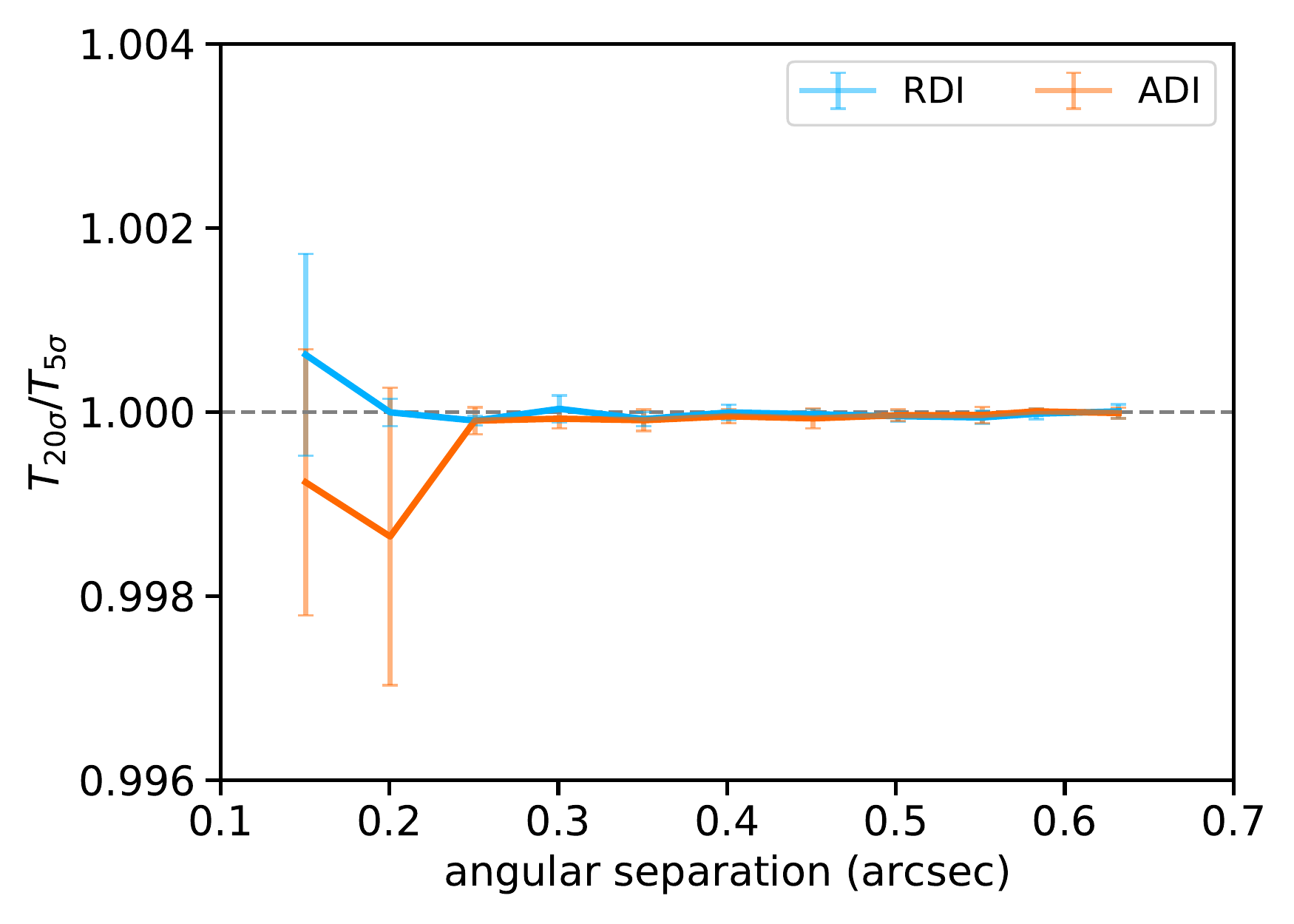}
      \caption{Ratios of throughput with different amounts of injected planet flux. The injected planet flux was scaled to be 20 or 5 times the noise. The throughputs of RDI and ADI are shown in blue and orange, respectively.
              }
          \label{Fig:validation_thoughput_estimation}
  \end{figure}

We estimated the throughput of the PCA subtraction using the simulated planet injection. To validate the amount of injected planet flux, we injected the planet flux at 20 or 5 times the noise. We tested this on the same target sample used in Sect.~\ref{subsect:impact_of_m_size}, eight targets in total. We estimated the average throughput and corresponding uncertainty by taking the mean and standard deviation of the results from the selected eight targets. Fig.~\ref{Fig:validation_thoughput_estimation} shows the difference in throughput with different amounts of injected planet flux. Neither RDI nor ADI show a difference in throughput with different injected fluxes at separations from 0.15\arcsec-0.65\arcsec.  
   \FloatBarrier
It is expected that different amounts of injected planet flux lead to almost the same throughput because the projection of the planet signal on PCs is a linear calculation \citep{Soummer2012}. Thus the small change in planet flux will not significantly change the throughput of the PCA subtraction. In the first step of PCA, we need to subtract the average values of reference and science images to ensure they have zero mean. However, the injection of simulated planets has a minor impact on the mean value of the image. This is because we only injected planet flux in a small fraction of the image (i.e., \textless1\%) compared to the entire FoV. The injected flux was also low, at the planetary rather than stellar level. 


\section{Selected samples}
Tables~\ref{table:sample_delta_Seeing} and \ref{table:sample_delta_PA} list the observations of targets in our selected samples (see Sect.~\ref{subsec:sample_selection} for details).\nopagebreak[4]
\begin{table*}
\caption{Targets with different seeing conditions.}             
\label{table:sample_delta_Seeing}      
\centering                          
\begin{tabular}{l c c c c c c c c c}        
\hline\hline                 
Name &  Program ID  & $n_{\rm DIT} \times t_{\rm DIT}^{(a)}$ & $t_{\rm tot}$ &  Seeing  & Field rotation & Windspeed & Strehl & $H$ & $F_{\rm AO}$ \\    
    &             &  &  (s) & (arcsec) & (degree) & (m s$^{-1}$)  &    & (mag) & (Hz) \\
    \hline                        
\object{HD\,207204}   &   097.C-1042   &   80$\times$32 &  2560  &  0.29   &   28.0  &  7.4   &  0.87  &  6.7   &   1380  \\
\object{HD\,133954}   &   0100.C-0543   &   144$\times$16 &  2304  &  0.30   &   27.2  &  7.9   &  0.78  &  7.7   &   1380  \\
\object{HD\,11506}   &   096.C-0241   &   36$\times$64 &  2304  &  0.37   &   28.1  &  6.4   &  0.82  &  6.3   &   1380  \\
\object{HD\,90884}   &   1100.C-0481   &   34$\times$96 &  3264  &  0.39   &   29.8  &  7.2   &  0.79  &  6.5   &   1380  \\
\object{*\,tau01\,Aqr}   &   095.C-0549   &   192$\times$8 &  1536  &  0.43   &   28.3  &  8.0   &  0.82  &  5.8   &   1200  \\
\object{V889\,Her}   &   198.C-0209   &   85$\times$64 &  5440  &  0.46   &   30.7  &  5.8   &  0.82  &  5.9   &   1380  \\
\object{LQ\,Hya}   &   097.C-0864   &   64$\times$32 &  2048  &  0.47   &   33.8  &  6.0   &  0.84  &  5.6   &   1380  \\
\object{HD\,97244}   &   1100.C-0481   &   48$\times$96 &  4608  &  0.51   &   27.4  &  7.8   &  0.79  &  5.8   &   1380  \\
\object{HD\,24966}   &   097.C-1042   &   75$\times$32 &  2400  &  0.51   &   34.5  &  5.7   &  0.85  &  6.9   &   1380  \\
\object{HD\,8558}   &   096.C-0241   &   64$\times$64 &  4096  &  0.53   &   28.2  &  6.0   &  0.66  &  6.9   &   1200  \\
\object{HD\,207575}   &   095.C-0298   &   80$\times$64 &  5120  &  0.54   &   32.2  &  5.9   &  0.85  &  6.1   &   1200  \\
\object{BD+20\,1790}   &   198.C-0209   &   80$\times$64 &  5120  &  0.58   &   27.3  &  7.4   &  0.70  &  7.0   &   1380  \\
\object{HD\,204277}   &   1100.C-0481   &   64$\times$96 &  6144  &  0.68   &   34.0  &  6.4   &  0.84  &  5.5   &   1380  \\
\object{HD\,156751}   &   1100.C-0481   &   48$\times$96 &  4608  &  0.69   &   31.3  &  8.4   &  0.81  &  6.3   &   1380  \\
\object{HD\,1466}   &   096.C-0241   &   64$\times$64 &  4096  &  0.69   &   25.0  &  8.3   &  0.62  &  6.2   &   1200  \\
\object{HD\,1466}   &   097.C-0865   &   80$\times$64 &  5120  &  0.69   &   31.1  &  5.9   &  0.81  &  6.2   &   1380  \\
\object{HD\,104125}   &   096.C-0241   &   64$\times$64 &  4096  &  0.73   &   28.4  &  6.0   &  0.78  &  6.3   &   1380  \\
\object{HD\,45270}   &   096.C-0241   &   256$\times$16 &  4096  &  0.91   &   28.4  &  6.8   &  0.79  &  5.2   &   1380  \\
\object{BD+21\,1764}   &   198.C-0209   &   70$\times$64 &  4480  &  1.0   &   23.7  &  5.7   &  0.69  &  6.2   &   1380  \\
\hline
\end{tabular}
\tablefoot{
\tablefoottext{a}{$n_{\rm DIT}$ is the number of image frames and $t_{\rm DIT}$ is exposure time per image frame.} The median values of seeing, PA rotation, wind speed, Strehl ratio, $H$ magnitude, and AO loop frequency ($F_{\rm AO}$) in our master reference library are 0.53\arcsec, 28.9$^{\circ}$, 7.1~m~s$^{-1}$, 0.75, 6.6 mag, and 1380 Hz. 
}
\end{table*}

\begin{table*}
\caption{Targets with different PA rotations.}             
\label{table:sample_delta_PA}      
\centering                          
\begin{tabular}{l c c c c c c c c c c}        
\hline\hline                 
Name &  Program ID  & $n_{\rm DIT} \times t_{\rm DIT}^{(a)}$ &  $t_{\rm tot}$ & Field rotation & Seeing  & Windspeed & Strehl & $H$ &  $F_{\rm AO}$  \\    
    &             &  &  (s) & (degree) & (arcsec) & (m s$^{-1}$)  &    & (mag) & (Hz) \\
    \hline                        
\object{HD\,81485B}   &   095.C-0346   &   48$\times$32 &  1536  &  8.0   &   0.46  &  7.6   &  0.75  &  7.6   &   1380  \\
\object{HD\,208233}   &   097.C-0826   &   144$\times$16 &  2304  &  13.7   &   0.59  &  5.9   &  0.62  &  6.9   &   1380  \\
\object{HD\,73267}   &   096.C-0241   &   16$\times$64 &  1024  &  14.3   &   0.48  &  8.0   &  0.72  &  7.1   &   1380  \\
\object{HD\,223340}   &   1100.C-0481   &   16$\times$96 &  1536  &  16.3   &   0.53  &  6.2   &  0.80  &  7.2   &   1380  \\
\object{HD\,16743}   &   097.C-1019   &   64$\times$32 &  2048  &  17.8   &   0.50  &  5.8   &  0.87  &  6.0   &   1380  \\
\object{HD\,77825}   &   198.C-0209   &   20$\times$64 &  1280  &  20.1   &   0.45  &  5.7   &  0.78  &  6.5   &   1380  \\
\object{HD\,105690}   &   098.C-0739   &   48$\times$64 &  3072  &  22.9   &   0.44  &  8.1   &  0.77  &  6.6   &   1380  \\
\object{BD+20\,1790}   &   198.C-0209   &   80$\times$64 &  5120  &  27.3   &   0.58  &  7.4   &  0.70  &  7.0   &   1380  \\
\object{HD\,97244}   &   1100.C-0481   &   48$\times$96 &  4608  &  27.4   &   0.51  &  7.8   &  0.79  &  5.8   &   1380  \\
\object{HD\,8558}   &   096.C-0241   &   64$\times$64 &  4096  &  28.2   &   0.53  &  6.0   &  0.66  &  6.9   &   1200  \\
\object{V889\,Her}   &   198.C-0209   &   85$\times$64 &  5440  &  30.7   &   0.46  &  5.8   &  0.82  &  5.9   &   1380  \\
\object{HD\,207575}   &   095.C-0298   &   80$\times$64 &  5120  &  32.2   &   0.54  &  5.9   &  0.85  &  6.1   &   1200  \\
\object{LQ\,Hya}   &   097.C-0864   &   64$\times$32 &  2048  &  33.8   &   0.47  &  6.0   &  0.84  &  5.6   &   1380  \\
\object{HD\,24966}   &   097.C-1042   &   75$\times$32 &  2400  &  34.5   &   0.51  &  5.7   &  0.85  &  6.9   &   1380  \\
\object{BD+01\,2063}   &   1100.C-0481   &   44$\times$96 &  4224  &  41.3   &   0.46  &  7.2   &  0.78  &  6.2   &   1380  \\
\object{HD\,75519}   &   198.C-0209   &   64$\times$64 &  4096  &  50.4   &   0.55  &  6.7   &  0.79  &  6.3   &   1380  \\
\object{HD\,212658}   &   1100.C-0481   &   48$\times$96 &  4608  &  51.3   &   0.58  &  6.2   &  0.70  &  6.6   &   1380  \\
\object{HD\,119152}   &   097.C-1042   &   144$\times$16 &  2304  &  53.7   &   0.56  &  8.2   &  0.78  &  6.8   &   1380  \\
\object{HD\,219246}   &   1100.C-0481   &   30$\times$96 &  2880  &  77.7   &   0.50  &  6.5   &  0.82  &  7.2   &   1380  \\
\object{HD\,199443}   &   097.C-0865   &   64$\times$64 &  4096  &  79.6   &   0.56  &  6.3   &  0.87  &  5.5   &   1380  \\
\hline
\end{tabular}
\tablefoot{
\tablefoottext{a}{$n_{\rm DIT}$ is the number of image frames and $t_{\rm DIT}$ is exposure time per image frame.} The median values of seeing, PA rotation, wind speed, Strehl ratio, $H$ magnitude, and AO loop frequency ($F_{\rm AO}$) in our master reference library are 0.53\arcsec, 28.9$^{\circ}$, 7.1~m~s$^{-1}$, 0.75, 6.6 mag, and 1380 Hz. 
}
\end{table*}
\FloatBarrier

\section{Comparisons of throughput and noise}\label{appendix:comparison_throughput_noise}
\begin{figure*}
  \centering
    \includegraphics[width=0.9\textwidth]{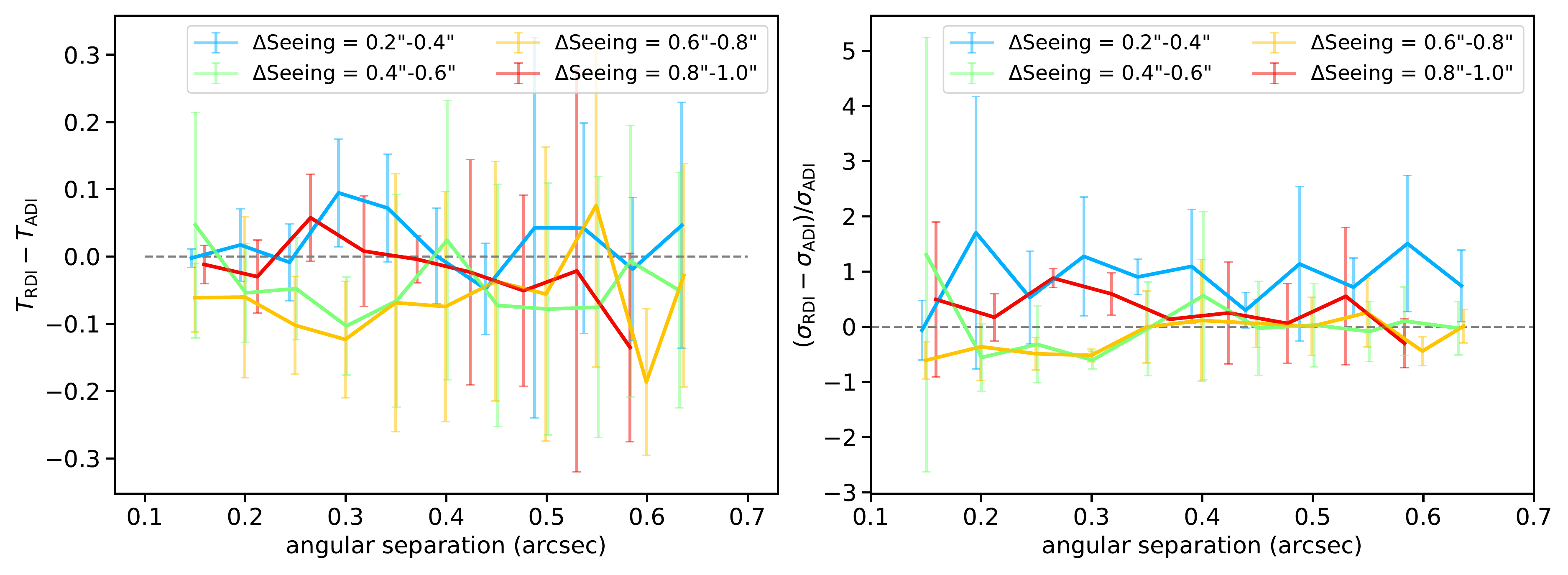}
      \caption{Differences in throughput (\textsl{left}) and noise (\textsl{right}) after the reductions of RDI and ADI. The absolute values of throughput and noise are shown in Fig.~\ref{Fig:RDI_vs_ADI_delta_Seeing_throughput_noise}
              }
           \label{Fig:comparison_TT_NN}
  \end{figure*}

The seeing condition affects the performances of RDI and ADI. To better compare the throughput and noise after the RDI and ADI reductions shown in Fig.~\ref{Fig:RDI_vs_ADI_delta_Seeing_throughput_noise}, we calculated their differences and present them in Fig.~\ref{Fig:comparison_TT_NN}. 
  \FloatBarrier


\section{Disk detections}
Table~\ref{table:disk_list} lists the detection of disks in this work with the parameters of the RDI reductions. 
\begin{table*}
\caption{Detected disks in this work.}             
\label{table:disk_list}      
\centering                          
\begin{tabular}{l c c c c c c c c}        
\hline\hline                 
Name & $H^{a}$  & Program ID  & Filed rotation &  ADI detection & RDI detection & PC mode$^{b}$ & Reference size$^{b}$\\    
 &  (mag) & & (degree) & & & & \\
\hline
\multicolumn{8}{c}{Planet-forming disks} \\
\hline
\object{V4046\,Sgr}   &   7.4   &   095.C-0298   &  83.8  &  Yes   &   Yes  &  100  &  500  \\
\object{PDS\,70}   &   8.8   &   095.C-0298   &  52.0  &  Yes   &   Yes  &  100  &  500  \\
\object{T\,Cha}   &   7.9   &   095.C-0298   &  28.9  &  Yes   &   Yes  &  50  &  500  \\
\object{RX\,J1615.3-3255}   &   8.8   &   095.C-0298   &  74.2  &  Yes   &   Yes  &  250  &  500  \\
\object{RY\,Lup}   &   7.7   &   097.C-0865   &  71.0  &  Yes   &   Yes  &  100  &  500  \\
\object{UX\,Tau\,A}   &   8.0   &   097.C-0865   &  16.4  &  No   &   Yes  &  40  &  200  \\
\object{V1094\,Sco}   &   9.0   &   099.C-0693   &  15.4  &  No   &   Yes  &  40  &  200  \\
\object{SZ\,Cha}   &   8.4   &   198.C-0209   &  24.7  &  Yes   &   Yes  &  100  &  500  \\
\object{J1604}   &   9.1   &   295.C-5034   &  91.9  &  Yes   &   Yes  &  100  &  500  \\
\object{HD\,34282}   &   8.5   &   096.C-0241   &  53.5  &  Yes   &   Yes  &  250  &  500  \\
\object{MWC\,758}   &   6.6   &   1100.C-0481   &  29.1  &  Yes   &   Yes  &  100  &  200  \\
\object{HD\,97048}   &   6.7   &   096.C-0241   &  24.5  &  Yes$^{c}$   &   Yes  &  100  &  500  \\
\object{HD\,100453}   &   6.4   &   096.C-0241   &  31.3  &  Yes   &   Yes  &  20  &  200  \\
\object{DG\,Tau\,A}   &   7.7   &   0104.C-0183   &  10.1  &  Yes   &   Yes  &  100  &  200  \\
\object{HD\,100546}   &   6.0   &   095.C-0298   &  34.7  &  Yes   &   Yes  &  100  &  500  \\
\object{TW\,Hya}   &   7.6   &   095.C-0298   &  76.7  &  No   &   Yes  &  20  &  200  \\
\object{RY\,Tau}   &   6.1   &   096.C-0241   &  19.5  &  Yes   &   Yes  &  40  &  200  \\
\object{HD\,169142}   &   6.9   &   095.C-0298   &  15.7  &  No   &   Yes  &  40  &  200  \\
\object{AK\,Sco}   &   7.1   &   097.C-0079   &  38.5  &  Yes   &   Yes  &  75  &  500  \\
\hline
\multicolumn{8}{c}{Debris disks} \\
\hline
\object{NZ\,Lup}   &   6.4   &   198.C-0209   &  60.4  &  Yes   &   Yes  &  100  &  500  \\
\object{HD\,141569}   &   6.9   &   095.C-0298   &  42.1  &  Yes   &   Yes  &  100  &  500  \\
\object{HD\,131835}   &   7.6   &   095.C-0298   &  72.6  &  Yes   &   Yes  &  100  &  500  \\
\object{HD\,106906}   &   6.8   &   095.C-0298   &  30.1  &  Yes   &   Yes  &  100  &  500  \\
\object{HR\,4796A}   &   5.8   &   1100.C-0481   &  64.1  &  Yes   &   Yes  &  100  &  500  \\
\object{HD\,129590}   &   7.9   &   097.C-0949   &  36.9  &  Yes   &   Yes  &  100  &  200  \\
\object{HD\,15115}   &   5.9   &   096.C-0241   &  29.6  &  Yes   &   Yes  &  250  &  500  \\
\object{HD\,120326}   &   7.6   &   097.C-0060   &  36.3  &  Yes   &   Yes  &  100  &  500  \\
\object{HD\,115600}   &   7.4   &   095.C-0298   &  27.3  &  Yes   &   Yes  &  100  &  500  \\
\object{HD\,111520}   &   7.8   &   097.C-0060   &  36.2  &  Yes   &   Yes  &  100  &  500  \\
\object{HD\,110058}   &   7.6   &   095.C-0607   &  10.7  &  Yes   &   Yes  &  100  &  500  \\
\object{HD\,131488}   &   7.8   &   0101.C-0753   &  28.9  &  Yes   &   Yes  &  100  &  500  \\
\object{GSC\,07396-00759}   &   8.8   &   198.C-0209   &  112.8  &  Yes   &   Yes  &  140  &  200  \\
\object{AU\,Mic}   &   4.8   &   095.C-0298   &  118.7  &  Yes   &   Yes  &  250  &  500  \\
\hline
\end{tabular}
\tablefoot{Here we show the data we used to make Fig.~\ref{Fig:RDI_disks}. We note that some targets may have multiple observations. For consistency in the paper, we fixed our FoV (\textless~0.735\arcsec). It is possible to increase the FoV to detect more extended disks using RDI. The ADI and RDI detections only indicate the detection within our adopted FoV. We note that this table should not be treated as a complete result of disk detections by SPHERE/IRDIS in the dual-band mode using the $H23$ filter. because our FoV is limited and we only included archival data released before 2021 January 1.
\tablefoottext{a}{The $H$-band magnitudes are adopted from \cite{2MASS_Skrutskie2006}}
\tablefoottext{b}{The parameters used in the RDI reduction. The number of PCs used in the PCA subtraction and the number of reference images used in the reference library.}
\tablefoottext{c}{No clear disk structure was detected in the ADI image using a FoV of 0.735\arcsec~in radius. However, we indeed detected similar disk structures as the ADI-PCA image reported in \cite{Ginski2016} when we adopted a larger FoV of 1.2\arcsec~in the ADI-PCA subtraction. }
}
\end{table*}

\end{appendix}

\end{document}